\documentclass[11pt]{article}
\usepackage{graphicx}

\newcommand{\BABARPubYear}    {04}

\input pubboard/babarsym

\long\def\inst#1{\par\nobreak\kern 4pt\nobreak
    {\it #1}\par\vskip 10pt plus 3pt minus 3pt}

\def\onlumi    {\ensuremath { 112.5  \invfb\ }}
\def\offlumi   {\ensuremath { 11.9 \invfb\  }}

\def\nBB   {\ensuremath {124.1 \times 10^{6}} }

\def\mes{\ensuremath{m_{\rm ES}}}

\def\de{\Delta E}
\def\qq{$q\bar{q}$}
\def\BB{$B\overline{B}$}
\def\Dm{\ensuremath{D^{-}}}
\def\Dzpi{\ensuremath{\kern 0.2em\overline{\kern -0.2em D}^{0}\pi^-}}
\def\Dzpiz{\ensuremath{\Dzb\piz}}
\def\Kpipi{\ensuremath{K^+ \pi^-\pi^-}}
\def\Kpi{\ensuremath{K^+ \pi^-}}

\def\BtoDstppbarpi {\ensuremath{B^{0} \to D^{*-} p \overline{p} \pi^{+}}}
\def\BtoDcppbarpi {\ensuremath{B^{0} \to D^{-} p \overline{p} \pi^{+}}}
\def\BtoDszppbar {\ensuremath{B^{0} \to \kern 0.2em\overline{\kern -0.2em D}^{*0} p \overline{p}}}
\def\BtoDzppbar {\ensuremath{B^{0} \to \kern 0.2em\overline{\kern -0.2em D}^{0} p \overline{p}}}

\def\Dstppbarpi {\ensuremath{D^{*-} p \overline{p} \pi^{+}}}
\def\Dcppbarpi {\ensuremath{D^{-} p \overline{p} \pi^{+}}}
\def\Dszppbar {\ensuremath{\kern 0.2em\overline{\kern -0.2em D}^{*0} p \overline{p}}}
\def\Dzppbar {\ensuremath{\kern 0.2em\overline{\kern -0.2em D}^{0} p \overline{p}}}

\def\bds   {\ensuremath{\Bz \to \Dstarm\proton\antiproton\pi^+}\xspace}
\def\bd   {\ensuremath{\Bz \to \Dm\proton\antiproton\pi^+}\xspace}
\def\bdsz {\ensuremath{\Bz \to \Dstarzb\proton\antiproton}\xspace}
\def\bdz   {\ensuremath{\Bz \to \Dzb\proton\antiproton}\xspace}

\def\bdsshort   {\ensuremath{\Bz \to \Dstarm\proton\antiproton\pi^+}}
\def\bdshort   {\ensuremath{,\Dm\proton\antiproton\pi^+}}
\def\bdszshort {\ensuremath{,\Dstarzb\proton\antiproton}}

\def\BR        {\mbox{$\mathcal{B}$}}

\begin{document}
{\pagestyle{empty}

\begin{flushright}
\babar-CONF-\BABARPubYear/20 \\
SLAC-PUB-10610 \\
August 2004 \\
\end{flushright}

\par\vskip 3cm

\begin{center}
\Large \bf \boldmath Measurement of the Branching Fractions for the
Decays \BtoDstppbarpi, \BtoDcppbarpi, \BtoDszppbar~and \BtoDzppbar
\end{center}
\bigskip

\begin{center}
\large The \babar\ Collaboration\\
\mbox{ }\\
\today
\end{center}
\bigskip \bigskip

\begin{center}
\large \bf Abstract
\end{center}
The $\Bz$ meson decay modes $\bdsshort\bdshort\bdszshort$, and $\bdz$
are studied in a sample of $124\times 10^{6}$ \BB~pairs collected with 
the $\babar$ detector at the PEP-II collider. 
The decay of $\bd$ is observed for the first time, with a measured 
branching fraction $\BR(\bd)=(3.80\pm
0.35\pm 0.46)\times 10^{-4}$. The following branching fractions are also
determined: $\BR(\BtoDstppbarpi)=(5.61\pm 0.59\pm 0.73)\times 10^{-4}$,
$\BR(\BtoDszppbar)=(0.67\pm 0.21\pm 0.09)\times 10^{-4}$,
and $\BR(\BtoDzppbar)=(1.24\pm 0.14\pm 0.12)\times 10^{-4}$. 
In each decay mode the invariant mass spectra of the charmed mesons and the
baryons are compared with a pure phase-space hypothesis 
in order to gain insight into the $B$ meson decay dynamics. In particular,
the Dalitz plots of $\Dstarzb\proton$
versus $\Dstarzb\antiproton$ for $\bdsz$ and of $\Dzb\proton$ versus
$\Dzb\antiproton$ for $\bdz$ are presented. All results are preliminary.

\vfill
\begin{center}

Submitted to the 32$^{\rm nd}$ International Conference on High-Energy Physics, ICHEP 04,\\
16 August---22 August 2004, Beijing, China

\end{center}

\vspace{1.0cm}
\begin{center}
{\em Stanford Linear Accelerator Center, Stanford University,
Stanford, CA 94309} \\ \vspace{0.1cm}\hrule\vspace{0.1cm}
Work supported in part by Department of Energy contract DE-AC03-76SF00515.
\end{center}

\newpage
} 

%
%
\begin{center}
\small

The \babar\ Collaboration,
\bigskip

%
B.~Aubert,
R.~Barate,
D.~Boutigny,
F.~Couderc,
J.-M.~Gaillard,
A.~Hicheur,
Y.~Karyotakis,
J.~P.~Lees,
V.~Tisserand,
A.~Zghiche
\inst{Laboratoire de Physique des Particules, F-74941 Annecy-le-Vieux, France }
% ...

\end{center}\newpage


\section{INTRODUCTION}
\label{sec:Introduction}

The very successful performance of the two \B factories PEP II and 
KEKB enables the study of \B meson decays with unprecedented 
sensitivity.  In this paper \B meson decays to final states 
which include a charmed meson and a baryon anti-baryon pair 
are studied. The observation of the $\BtoDstppbarpi$  
and $\ensuremath{B^{0} \to D^{*-} p \overline{n}}$ 
decays by CLEO~\cite{prl_86_2732}, and 
the color-suppressed $\BtoDzppbar$ and $\BtoDszppbar$ 
decay modes by Belle~\cite{prl_89_151802} suggest the dominance of 
multi-body final states in decays of \B mesons into baryons~\cite{cheng}.
In this paper we present the measurements of the branching fractions for the
following four decay modes: $\BtoDstppbarpi$, $\BtoDcppbarpi$,
$\BtoDszppbar$, and $\BtoDzppbar$. Observation of modes proposed here would
help clarify the dynamics of weak decays of \B involving baryons~\cite{Hou}.

Since the branching fractions of multi-body decays are
sizable~\cite{dunietz}, it is natural to ask whether such final states are
actually the products of an intermediate two-body channel.
If this is the case, then these initial two-body decays may involve a baryon-antibaryon bound
states~($N\overline{N}$)~\cite{yang, rosner} or a charmed pentaquark ($DN$, which
can be $\overline{c}uudd$ for example)~\cite{jaffe, wu} or a (nonexotic) heavy charmed
baryon. Motivated by these considerations, 
the invariant mass spectrum of the baryon-antibaryon and the invariant mass
spectra of the charmed meson and baryon are investigated.
In particular, the Dalitz plots for $\Dstarzb p$ versus
$\Dstarzb \overline{p}$ and $\Dzb p$ versus
$\Dzb \overline{p}$ for the $\BtoDzppbar$ and $\BtoDszppbar$ decay modes,
respectively, are presented. 
The inclusion of the charge conjugate modes is implicit throughout this report.


\section{THE \babar\ DETECTOR AND DATASET}
\label{sec:babar}
The data used in this analysis were collected with the \babar\ detector
at the \pep2\ storage ring. 
The sample corresponds to an integrated
luminosity of \onlumi at the \FourS\ resonance (on-resonance) 
and \offlumi taken $40\mev$ below the \FourS\ resonance 
(off-resonance). The on-resonance sample contains
about \nBB \xspace \BB\ pairs. The collider is operated with asymmetric
beam energies, producing a boost of $\beta\gamma \approx 0.56$ 
of the \FourS along the collision axis.

The \babar\ detector is optimized for asymmetric energy collisions at 
a center-of-mass (CM) energy corresponding to the \FourS\ resonance.
The detector is described in detail in reference~\cite{ref:babar}. 
Charged particle tracking is provided by a five-layer double-sided silicon 
vertex tracker (SVT) and a 40-layer drift chamber (DCH) contained 
within the magnetic field of a 1.5T superconducting solenoid.
The tracking system provides momentum reconstruction of charged
particles and measures energy-loss ($dE/dx$) for particle identification.
Additional charged $K$--$\pi$ particle identification is provided
by a ring-imaging Cherenkov detector (DIRC), which exploits the 
total internal reflection of Cherenkov photons within  
synthetic quartz bars. The energies of neutral particles are 
measured by an electromagnetic calorimeter (EMC) composed of 
6580 CsI(Tl) crystals. 
The magnetic flux return of
the solenoid (IFR) is instrumented with resistive plate chambers in order
to provide muon and neutral hadron identification.

A GEANT4-based \cite{geant4} Monte Carlo (MC) 
simulation is used to model the signal
efficiency and the physics backgrounds. Simulation samples
equivalent to approximately three times the accumulated data  were
used to model \BB\ events, and samples equivalent to approximately
one times the accumulated data were used to model $\epem \to$
\uubar, \ddbar, \ssbar, and \ccbar\ events.


\section{ANALYSIS METHOD}
\label{sec:Analysis}

The $\Bz$ meson is reconstructed in the following four decay modes:
$\BtoDstppbarpi$, $\BtoDcppbarpi$, $\BtoDszppbar$, and $\BtoDzppbar$.
The $\Dstarm$ and $\Dstarzb$ are reconstructed as $\Dzpi$ and
$\Dzpiz$, respectively. The $\Dm$ candidates are 
reconstructed in the decay mode $\Kpipi$. The $\Dzb$ candidates 
are reconstructed in the $\Kpi$ mode.

\subsection{$\Bz$ SELECTION}
\label{sec:BReco}

The $\Bz$ reconstruction proceeds as follows. First we reconstruct the
$\Dm$ and $\Dzb$ candidates in the decay modes noted above.
For all decay modes, charged kaons are distinguished 
from the pions and protons with energy-loss ($dE/dx$) information in the tracking system and
the Cherenkov angle and the number of photons measured by the DIRC.
For the $\BtoDszppbar$ decay mode only, the
pion from $\Dzb \to \Kpi$ decay must not be identified as either 
an electron, muon, kaon or proton.  The mass of the reconstructed $\Dzb$ or $\Dm$ 
candidates must be within 3 standard deviations of the fitted mean of
reconstructed $\Dzb$ or $\Dm$ mass for each decay mode. Finally, the
daughter tracks from the $\Dzb$ or $\Dm$ are required to be consistent with 
originating from a common vertex.
 
For the decay modes $\BtoDstppbarpi$ and $\BtoDszppbar$, $\Dstarm$ and
$\Dstarzb$ candidates are reconstructed by combining the $\Dzb$ 
candidate with a soft track or a soft $\piz$, respectively.
The momentum of the soft track or soft $\piz$ 
in the CM frame must be less than 0.45 \gev/c.
The $\piz$ candidate is required to have a two-photon invariant mass between
0.116--0.150 \gev/$c^2$, and its daughter photon candidates must
have a minimum cluster energy of 30 \mev.  
The mass difference between $D^{*}$ and $D^{0}$ ($\Delta M
=m_{D^{*}}-m_{D^{0}}$) is required to
be less than $0.1486~\gev/ c^2$ and $0.1453~\gev/c^2$ for $\Dstarm$ and
$\Dstarzb$ candidates, respectively. The $\Delta M$ requirement removes 
crossfeed from $\BtoDszppbar$ into $\BtoDzppbar$.   
For each track except for the soft pion from $\Dstarm$ decays, the
transverse momentum $p_T$ must be larger than 0.1~\gev/c in order to improve
the quality of the vertex fit.

To reconstruct the candidate $\BtoDstppbarpi$ and $\BtoDcppbarpi$ decay modes,
both proton and anti-proton candidates are distinguished
from pions and kaons on the basis of energy-loss ($dE/dx$) information in the
tracking system and the Cherenkov angle and the number of photons measured by
DIRC. The reconstructed $\Dstarm$ and $\Dm$, respectively, 
are combined with an identified proton and anti-proton pair and a track.
We require that the pion candidate track 
must not be identified as either an electron, a muon, 
a kaon or a proton. For the decay modes $\BtoDszppbar$ and 
$\BtoDzppbar$, the $\Dstarzb$ and $\Dzb$ candidates, respectively,
are combined with an identified 
proton and anti-proton pair. All the daughters of the $\Bz$ must originate from a common vertex.

Two additional kinematic variables are used to identify the reconstructed $\Bz$
candidates~\cite{ref:babar}. The first is the beam-energy-substituted mass, 
$m_{ES}=[(E^2_{CM}/2+\mbox{\bf p}_i\cdot \mbox{\bf p}_B)^2/E^2_i-{\mbox{\bf p}^2_B}]^{1/2}$,
where $E_{CM}$ is the total CM energy of the $e^+e^-$ collision,
($E_i, \mbox{\bf p}_i$) is
the four-momentum of the initial  $e^+e^-$ system 
and ${\bf p}_B$ is the momentum of the 
reconstructed $B$ candidate, both measured in the laboratory frame.
The second variable is $\Delta E=E^*_B-E_{CM}/2$, where
$E^*_B$ is the $B$-candidate energy in the CM frame.
The \B meson candidates are defined by requiring:
$5.20 \gev/c^2 <\mes<5.29\,\gev/c^2$ and $\left|\DeltaE\right|<0.12\,\gev$.
These two variables are then used in a maximum likelihood fit to 
extract signal and background yields. 

If more than one suitable $\Bz$ candidate is
reconstructed in an event, then for each $\Bz$ decay mode one
best candidate is selected.  
The best candidate selection algorithm is based
on a $\chi^2$-like quantity constructed from the difference between the $D$
mass and/or mass difference $\Delta M$ for the candidate and the nominal
value~\cite{ref:pdg2002}. For a given event,
the candidate with the lowest value of  $\chi^2$ is selected for each $\B^0$
decay mode. For the $\BtoDstppbarpi$ and 
$\BtoDszppbar$ decays, the best candidate is selected based on
the mass of the $\Dzb$ candidate and $\Delta M$. For the $\BtoDcppbarpi$
and $\BtoDzppbar$ decay modes, the reconstructed $D$ meson
mass is used for best candidate selection.

To suppress background from two-jet-like $\epem \to$
\qq~continuum, variables that characterize the event topology are
used. We require $\cos{\theta_{thr}}<0.9$, where $\theta_{thr}$ is the angle
between the thrust axis of the $\Bz$ candidate and that of the 
rest of the event. This requirement eliminates 63\% of the 
continuum background and retains 88\% of the signal events.
For further continuum background suppression we 
require that the ratio of the second to zeroth
Fox-Wolfram moments~\cite{r2definision} is less than 0.35.

The signal efficiency in each $B$ decay mode
after applying all selection criteria is shown in
Table~\ref{tab:effi-yields}. The efficiencies listed in the 
table are obtained from signal MC simulation assuming pure phase-space for
the \B decay model.

\begin{table}[h]
\begin{center}
\caption{ Summary of selection efficiencies obtained from MC simulation
based on a phase-space
model and fitted yields from data for the \B decay modes in this report. 
The error on the efficiency
comes from the Monte Carlo statistics. The statistical significance is
calculated as $\sqrt{2\Delta \log{\cal L}}$, where $\Delta \log{\cal L}$ is
the log-likelihood difference between a signal hypothesis corresponding to
the yield and that corresponding to a null yield.}
\label{tab:effi-yields}
\begin{tabular}{lccc}
\hline\hline
 Mode              & Efficiency, \% & Signal yield    & Statistical significance \\ \hline
 \BtoDstppbarpi    & 6.97 $\pm$ 0.10 & 130$\pm$14 &  18$\sigma$\\
 \BtoDcppbarpi     & 5.87 $\pm$ 0.10 & 238$\pm$22 &  17$\sigma$\\
 \BtoDszppbar      & 5.53 $\pm$ 0.09 & 13$\pm$ 4  &  5$\sigma$\\
 \BtoDzppbar       & 16.53$\pm$ 0.15 & 96 $\pm$ 11 & 17$\sigma$ \\ \hline \hline
\end{tabular}
\end{center}
\end{table}

\subsection{THE MAXIMUM LIKELIHOOD FIT}
\label{sec:MLfit}

We perform an unbinned extended maximum
likelihood (ML) fit to extract the signal and background yields. 
The variables $\mes$, $\de$ are
used to separate signal from backgrounds.  The data sample is assumed to consist of signal
and combinatorial backgrounds,  which arise from random combinations of
charged or neutral candidates from
both continuum and \BB~events. The extended likelihood for a sample of $N$
candidates is 
\begin{equation}
\mathcal{L}=e^{-N^{\prime}}\cdot\prod_{i=1}^{N}\lbrace
N_{sig}\cdot {\cal P}_{sig}(m_{ESi},\Delta E_i)
+N_{bkg}\cdot {\cal P}_{bkg}(m_{ESi},\Delta E_{i})\rbrace
\label{eqn:likelihood}
\end{equation}
where $N_{sig}$ and $N_{bkg}$ are the number of signal and
background yields (to be determined by the fit), respectively, and
$N^{\prime} = N_{sig}$+$N_{bkg}$.
The probability density functions (PDFs) ${\cal P}_{sig}$ and ${\cal P}_{bkg}$ 
are the product of the PDFs of two discriminating variables. The signal PDF
is thus given by ${\cal P}_{sig} = {\cal P}(m_{ESi})\cdot{\cal P}(\Delta
E_{i})$.  The signal PDFs are decomposed into two parts with distinct
distributions:
\begin{itemize}
\item Signal class I: signal events that are correctly reconstructed or
      signal events that are misreconstructed
      due to a random slow $\pi^-$ or $\pi^0$ assigned to a $D^{*-}$ or $D^{*0}$
      decay.  Since the $\mes$ and $\Delta E$ distributions for these two kinds
      of events do not significantly differ from each other, we put them into
      one signal class in the ML fit. 
\item Signal class II: signal events that are misreconstructed due to wrong tracks from \B
      direct decays or from a wrongly reconstructed $D$ meson. The $\mes$ and
      $\Delta E$ distributions are significantly different from those of signal class I.
      Hereafter signal class II
      events are also called Self-Cross-Feed (SCF) events. The fraction of
      this kind of misreconstructed
      signal event is estimated by MC simulation.
\end{itemize}
The PDFs of $\mes$ and $\Delta E$ for signal events are taken from the MC simulation, 
with the exception that the means of the signal Gaussian for the $\mes$ and $\Delta E$ PDFs
are free to vary in the fit. 
The $m_{ES}$ and $\Delta E$ PDFs for the combinatorial backgrounds, which
include \B background and continuum background, are described by two free
parameters.  One of these is from an ARGUS function~\cite{argus} used to
describe the $\mes$ shape, the other is
from a first-order polynomial for the $\Delta E$ shape. A total of six parameters, including
signal and background yields and all the parameters related to background
PDFs,  are varied in the fit.


\section{PHYSICS RESULTS}
\label{sec:Physics}

\begin{figure}[bp]
\begin{center}
\resizebox{12cm}{!}{
        \includegraphics{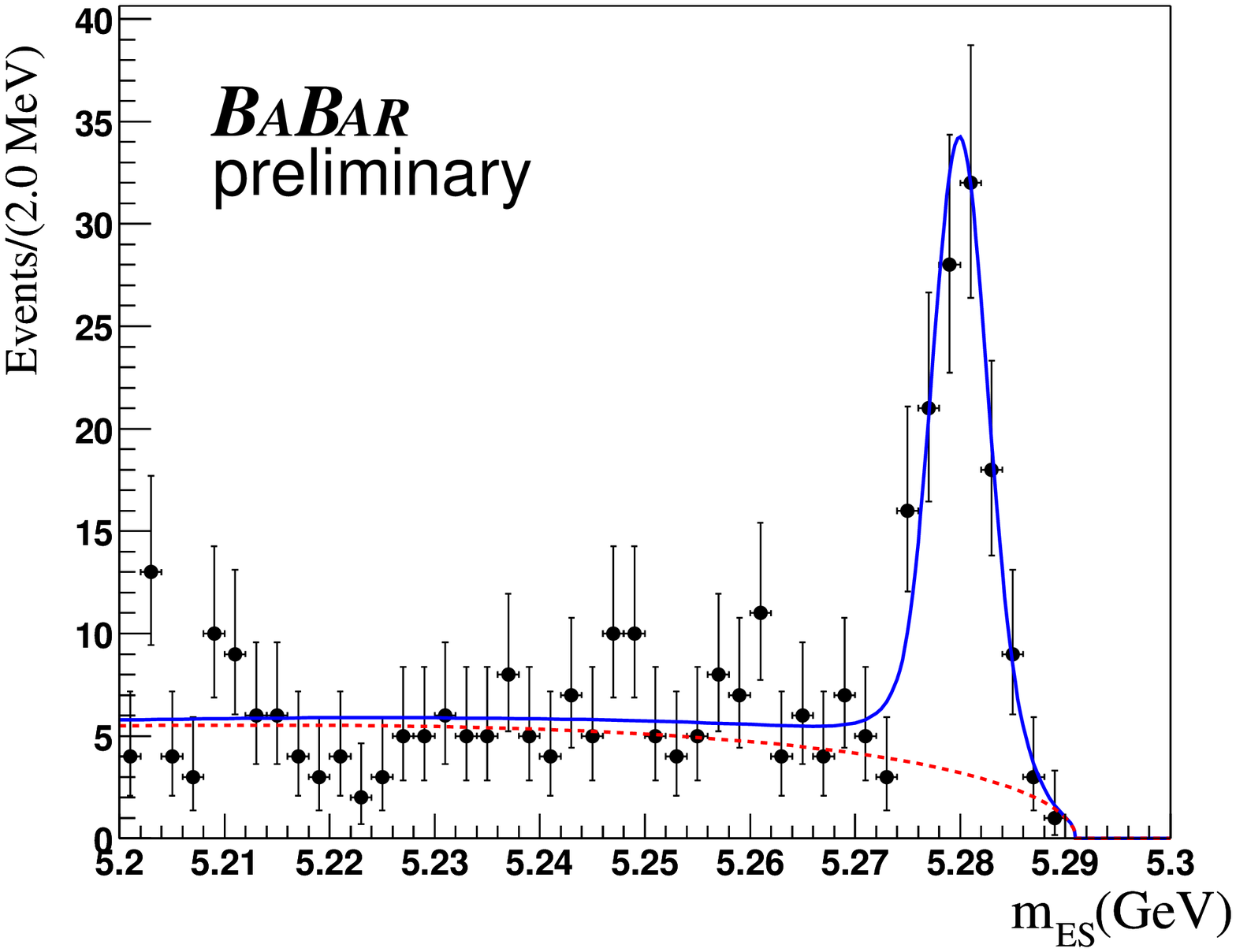}\hspace{0.3cm}
        \includegraphics{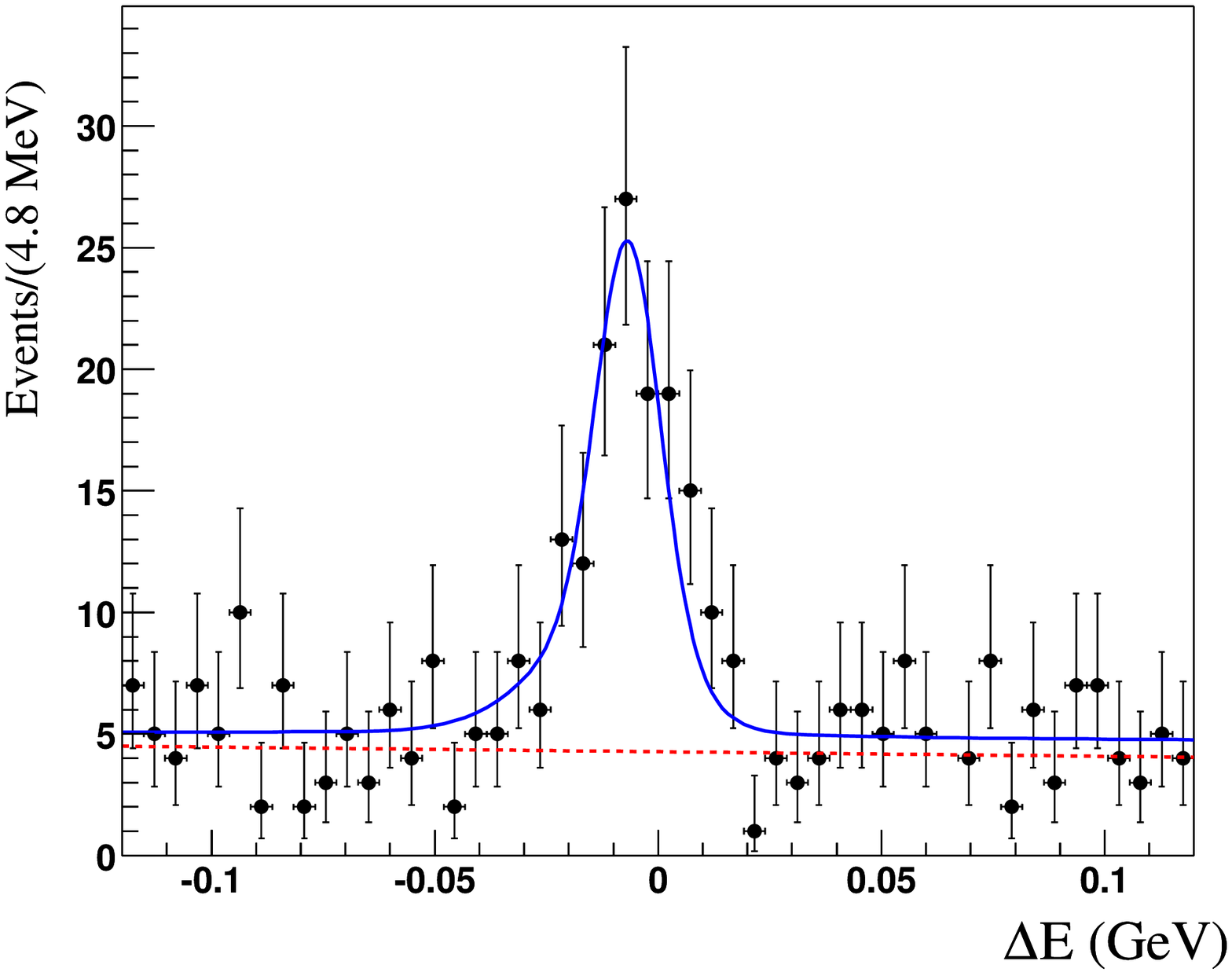}
}
\resizebox{12cm}{!}{
        \includegraphics{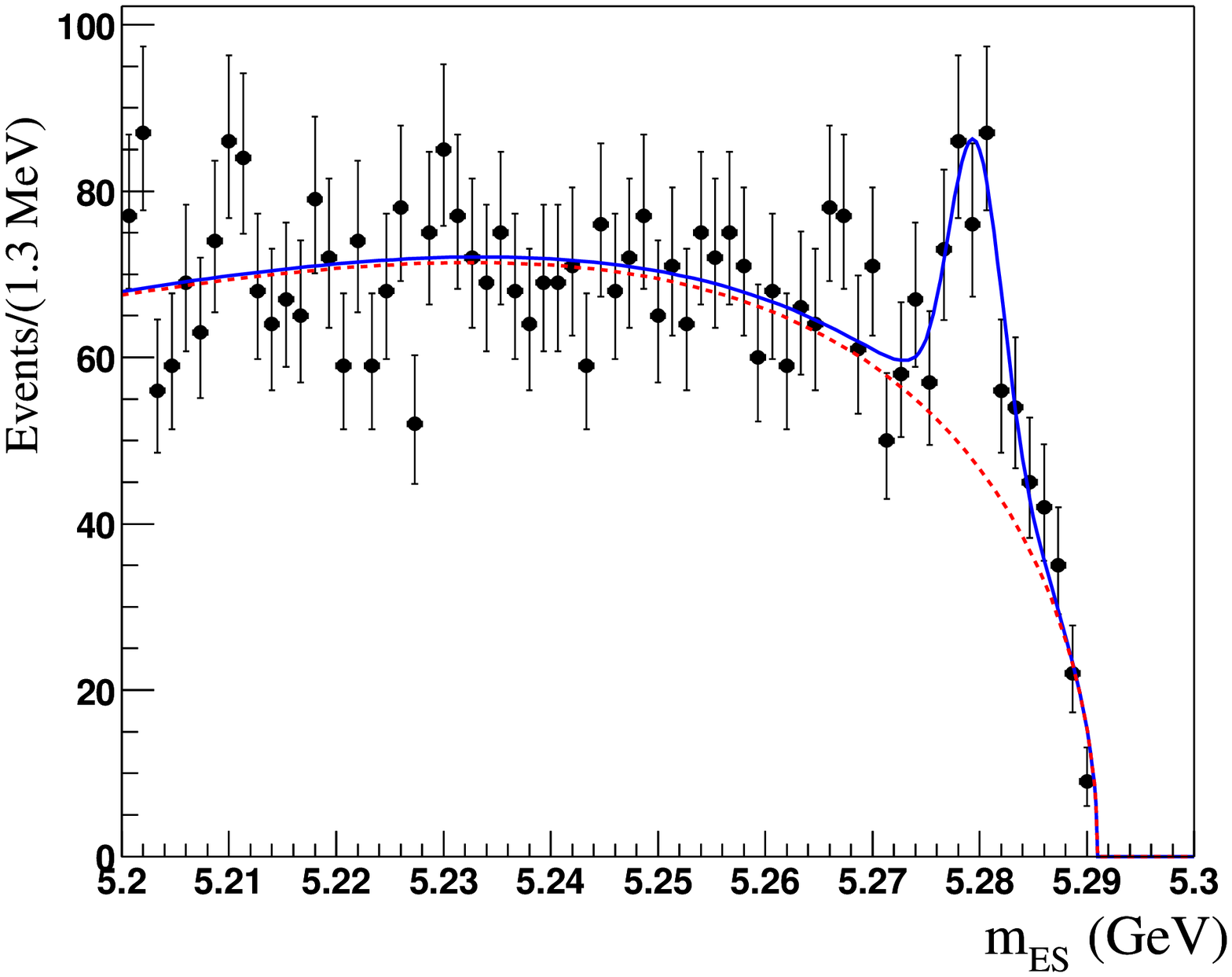}\hspace{0.3cm}
        \includegraphics{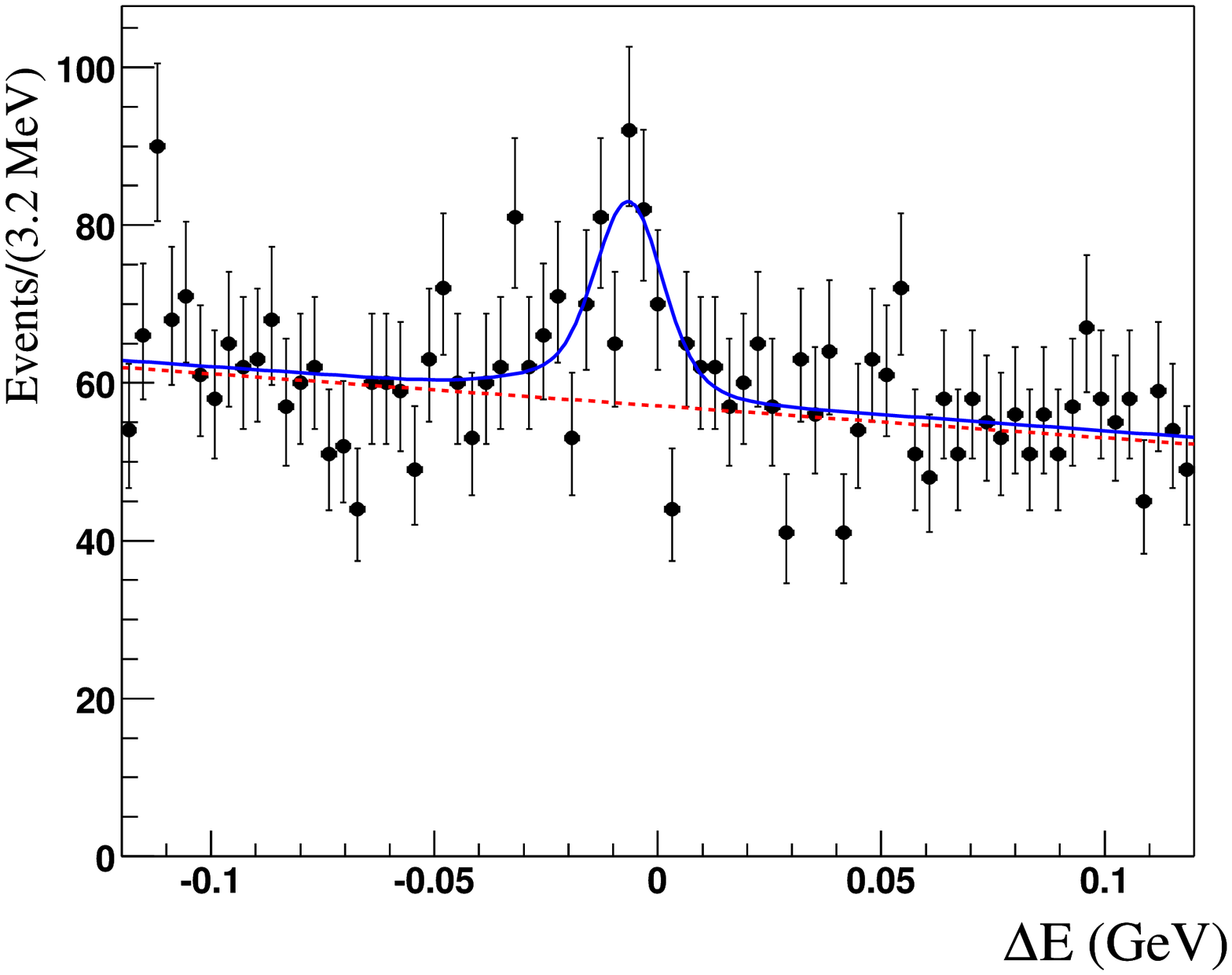}
}
\resizebox{12cm}{!}{
        \includegraphics{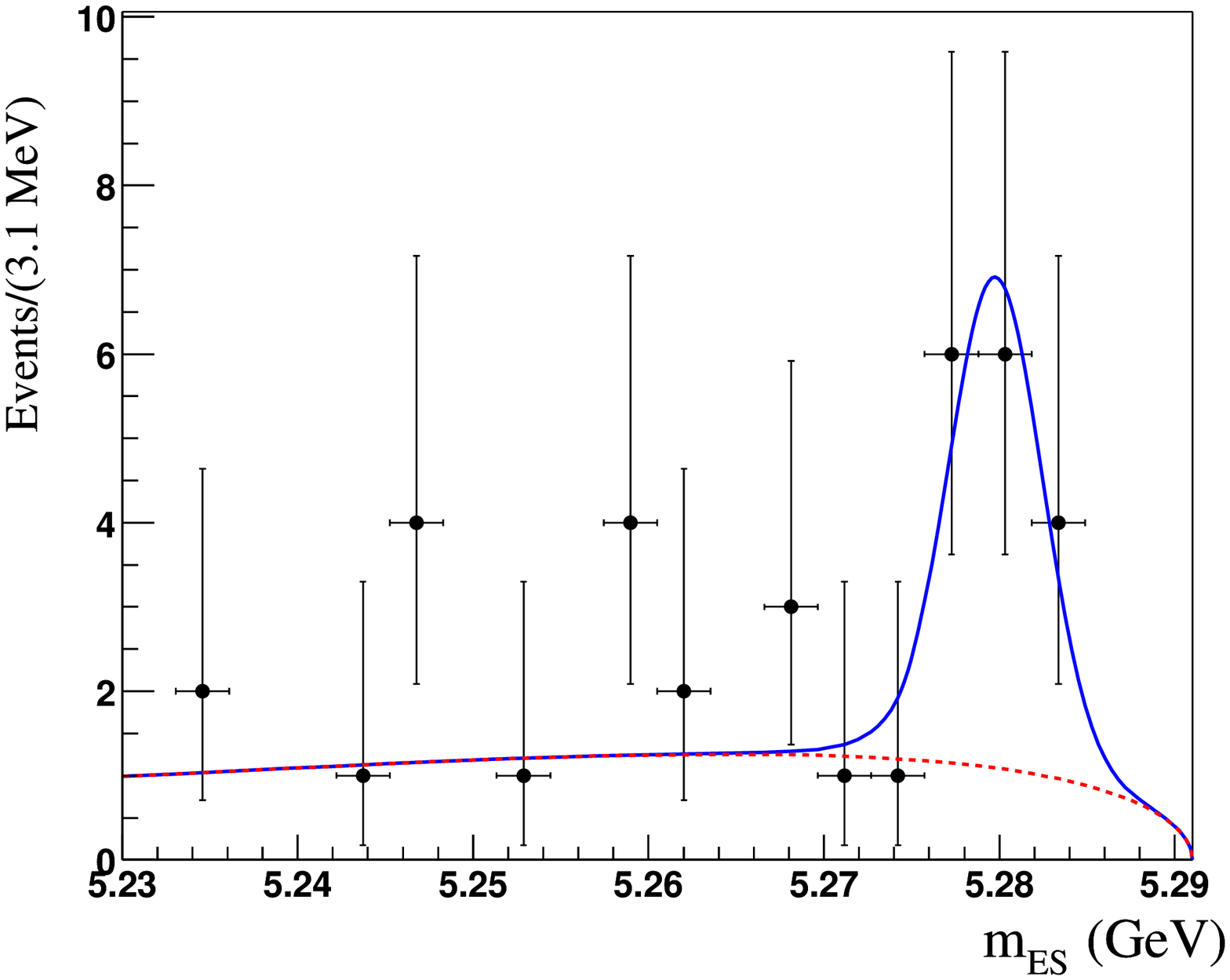}\hspace{0.3cm}
        \includegraphics{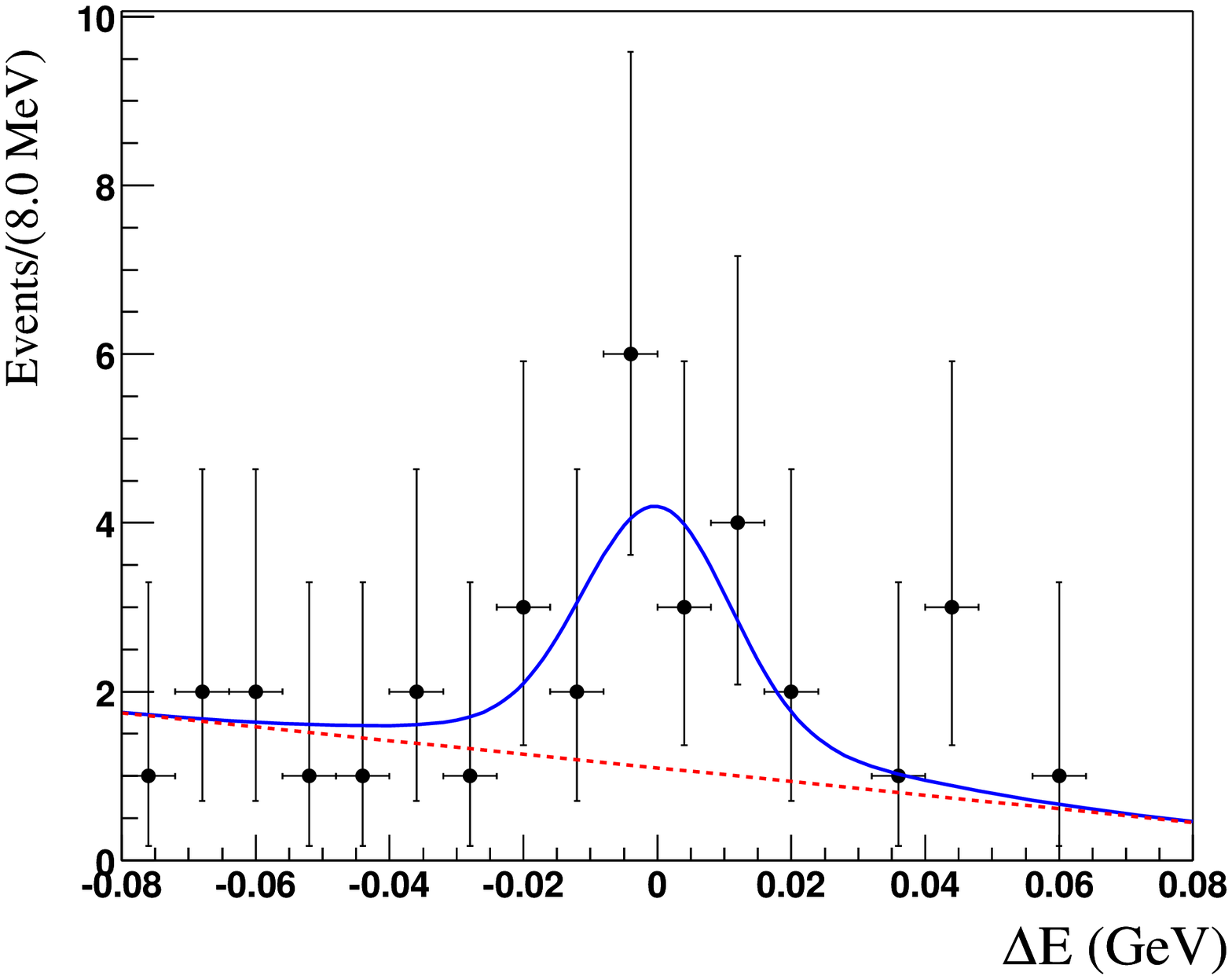}
}
\resizebox{12cm}{!}{
        \includegraphics{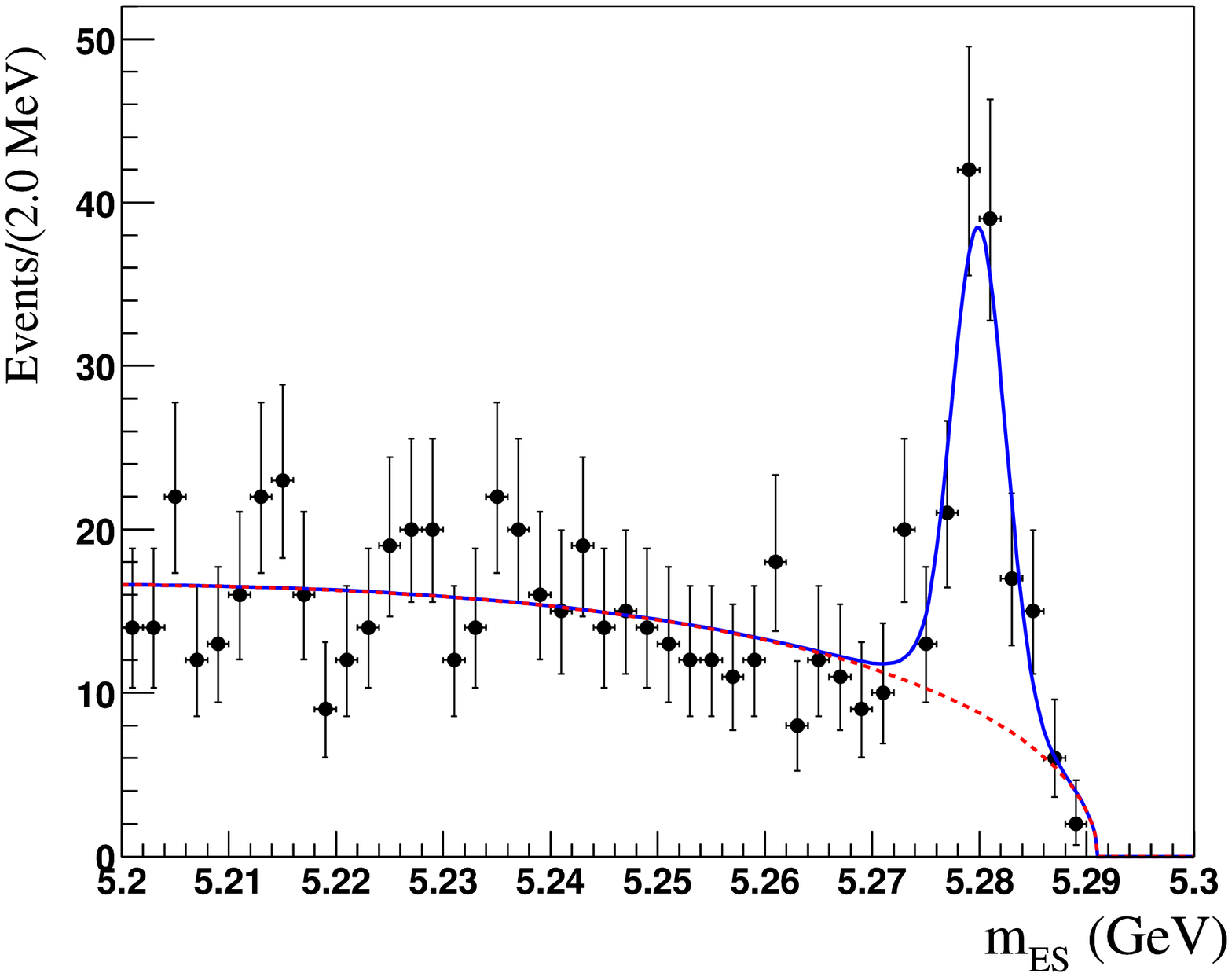}\hspace{0.3cm}
        \includegraphics{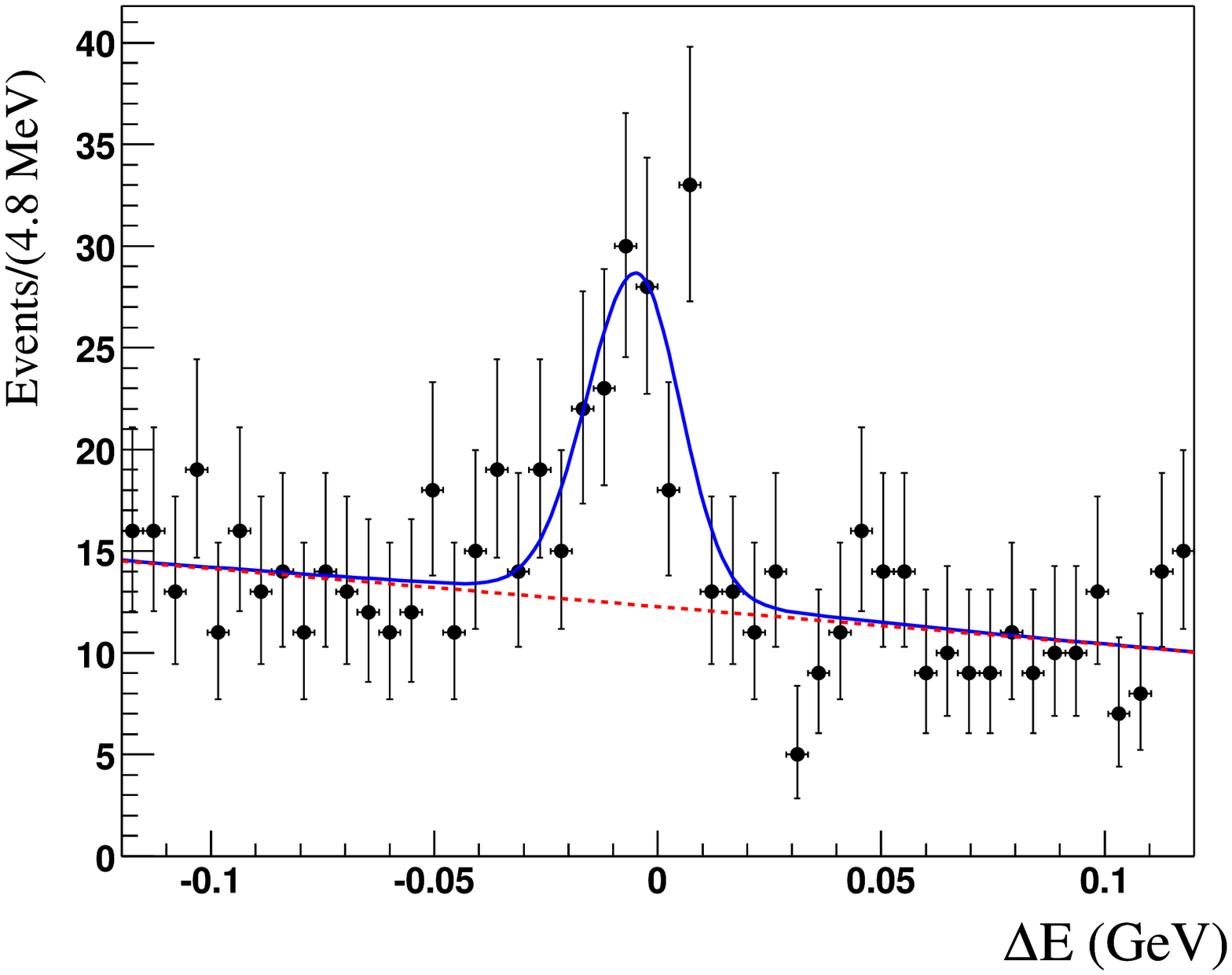}
}

\caption {$\mes$ (left-hand column) and $\Delta E $ (right-hand column) distributions
in data. The plots from top to bottom are for $\bds$, $\bd$, $\bdsz$,
and $\bdz$, respectively. The solid curve represents a projection of the
maximum likelihood fit. The dashed curve represents the contribution from combinatorial
background.}
\label{fig:proj}
\end{center}
\end{figure}
\begin{figure}[bp]
\begin{center}
\resizebox{12cm}{!}{
        \includegraphics{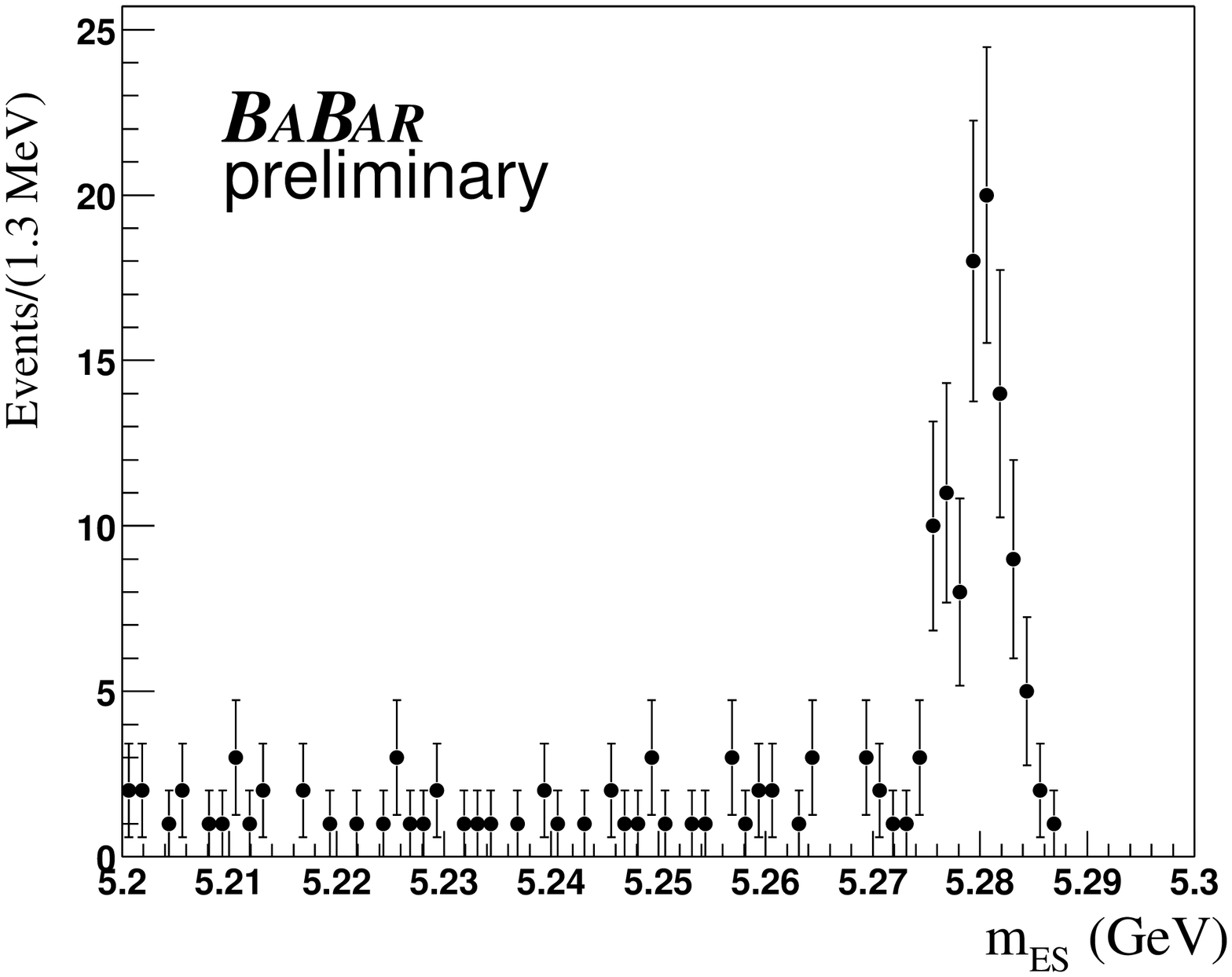}\hspace{0.3cm}
        \includegraphics{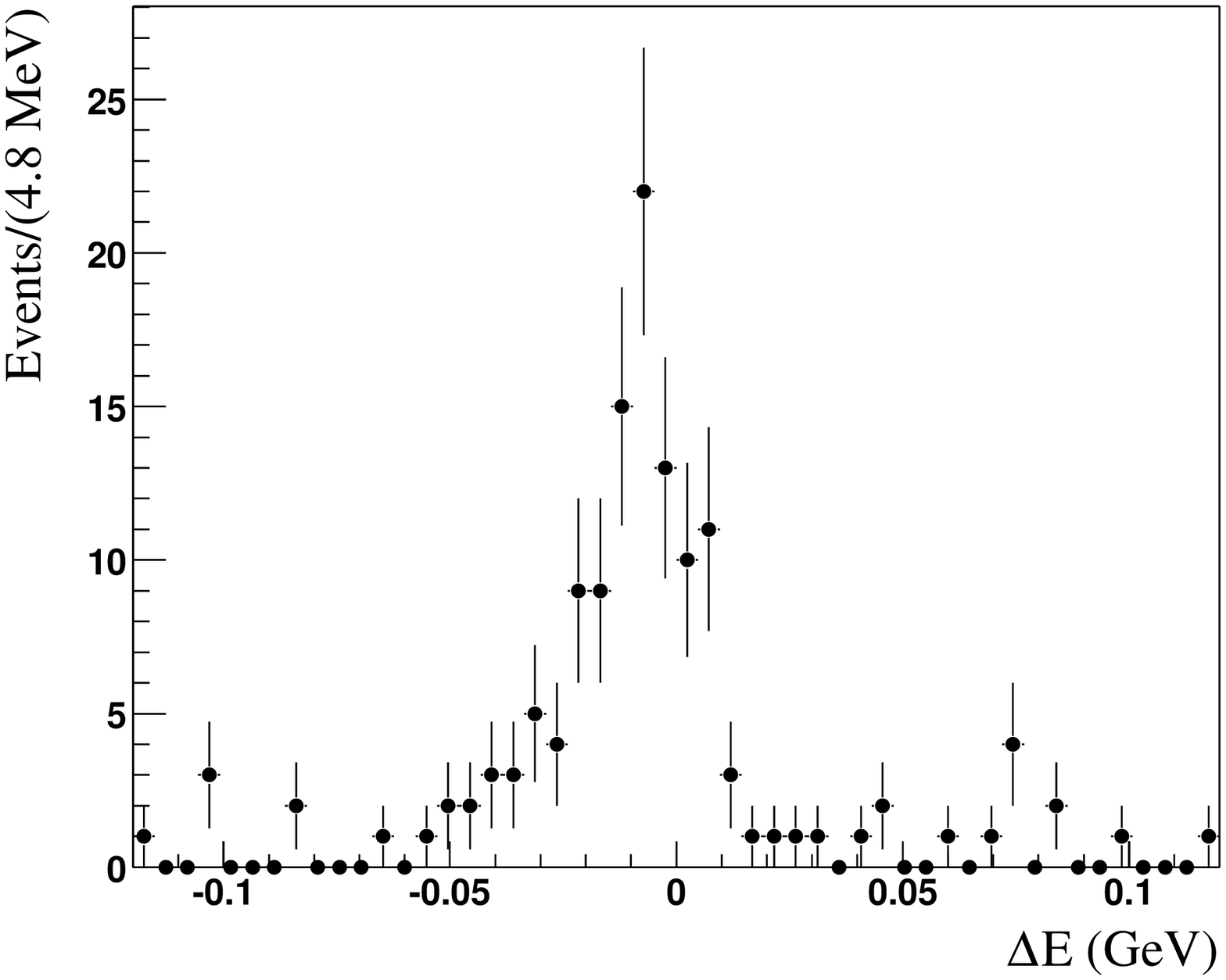}
}
\resizebox{12cm}{!}{
        \includegraphics{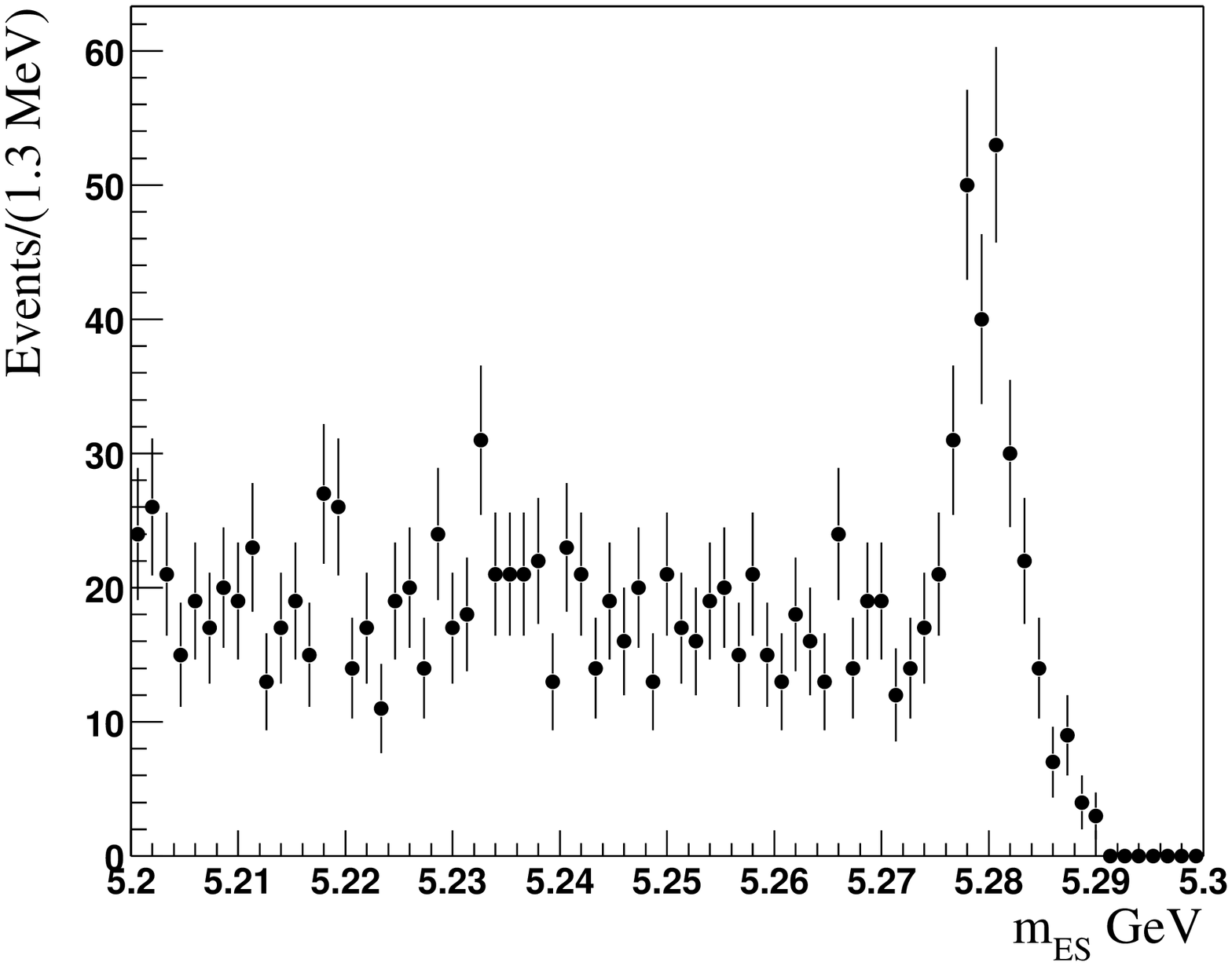}\hspace{0.3cm}
        \includegraphics{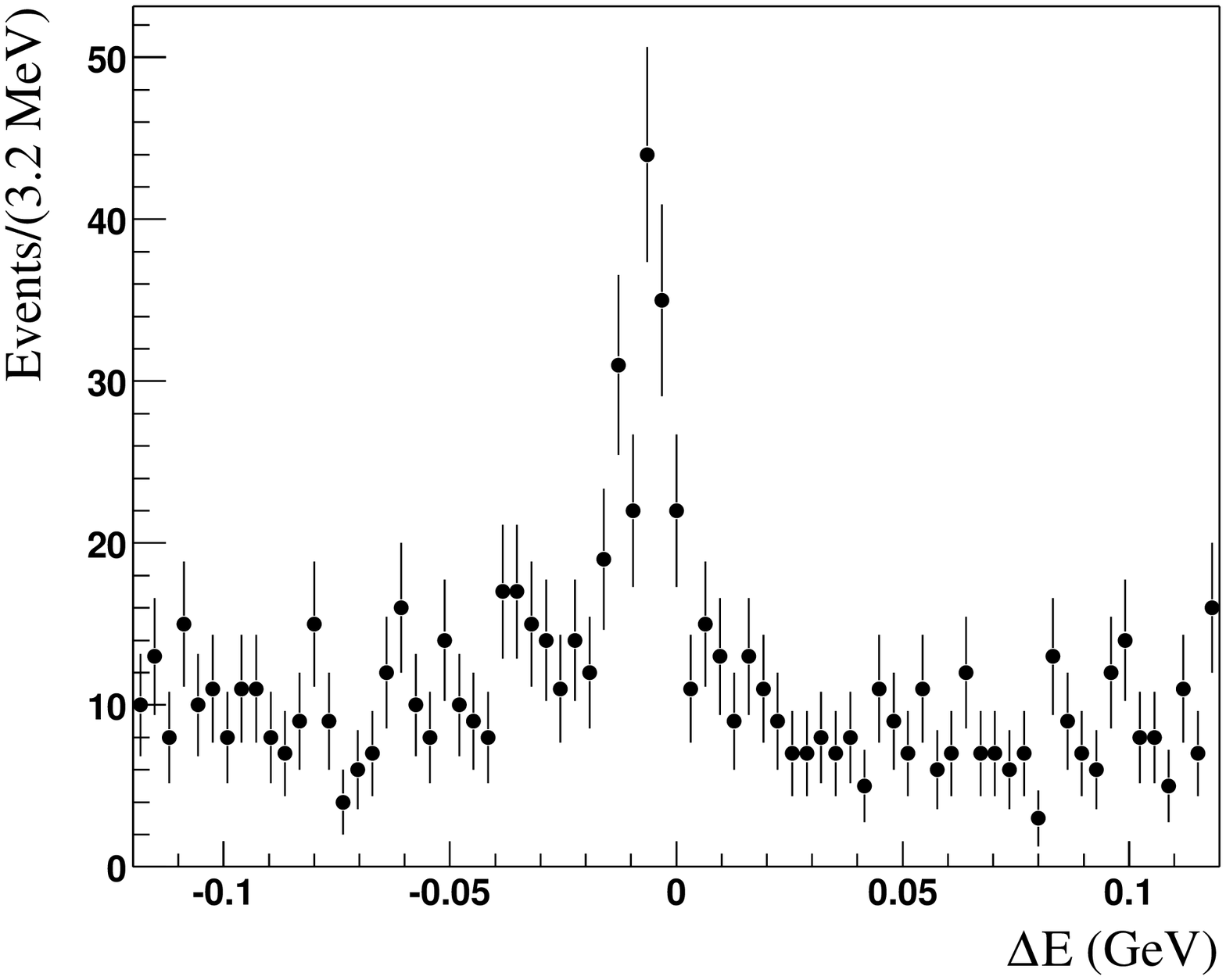}
}
\resizebox{12cm}{!}{
        \includegraphics{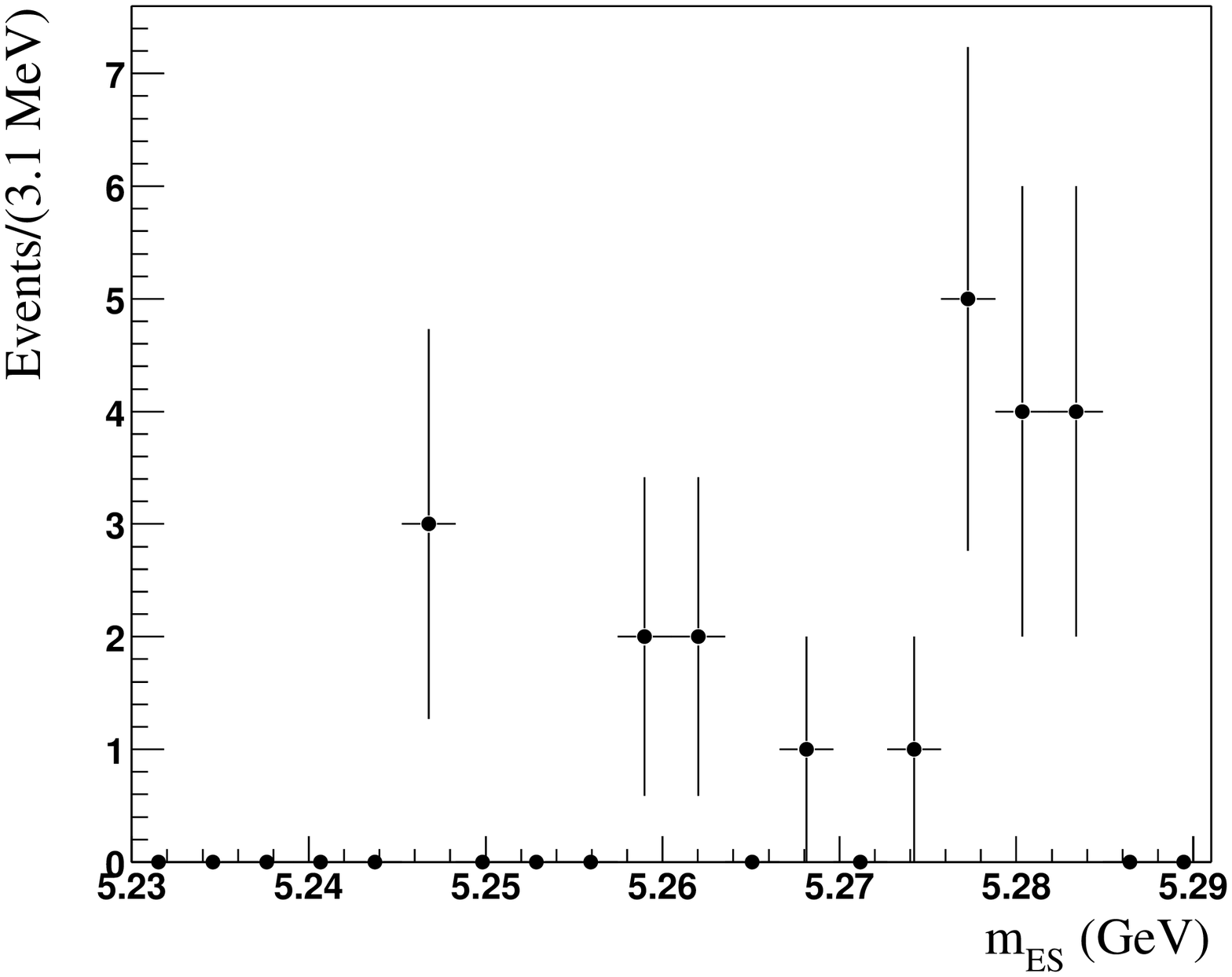}\hspace{0.3cm}
        \includegraphics{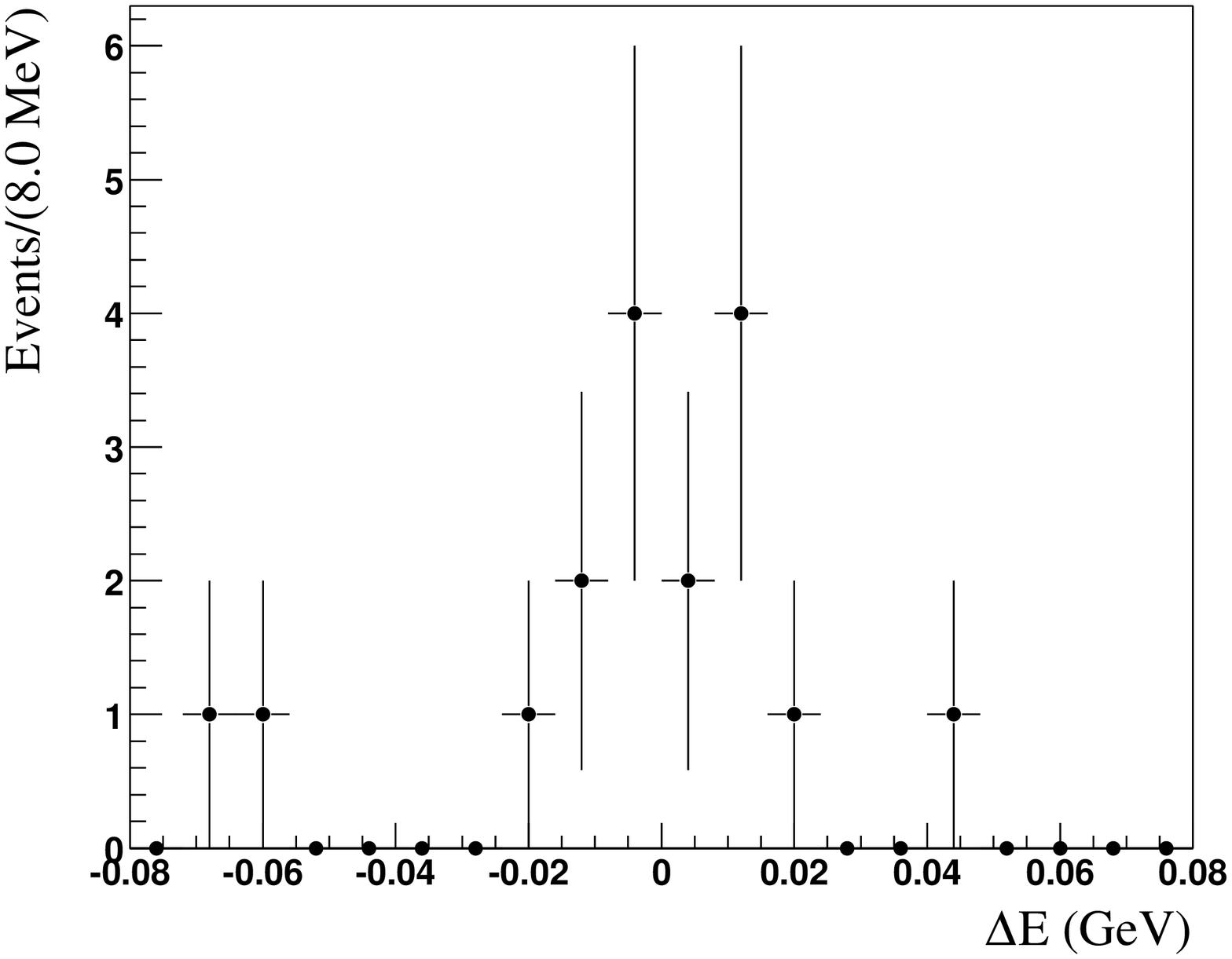}
}
\resizebox{12cm}{!}{
        \includegraphics{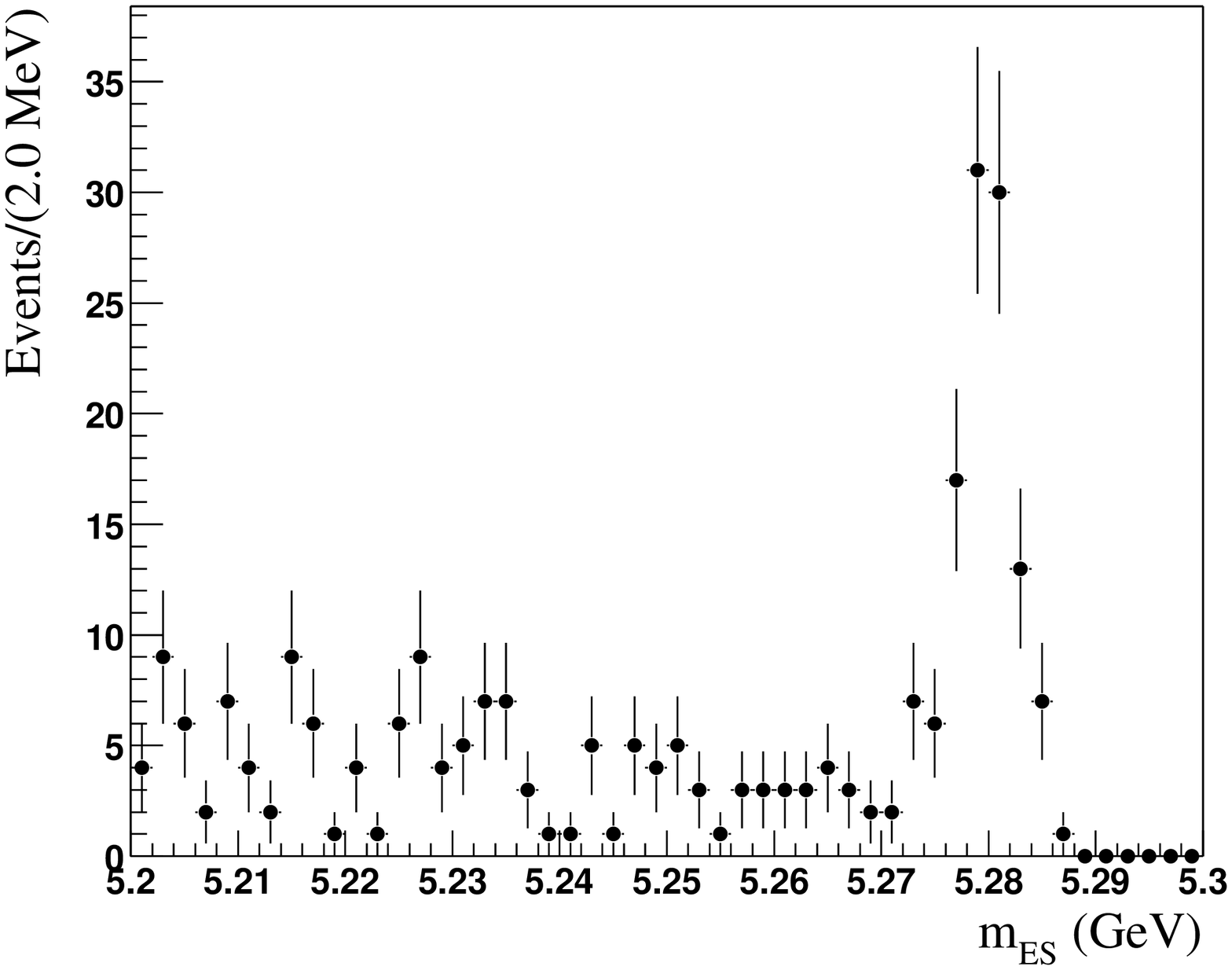}\hspace{0.3cm}
        \includegraphics{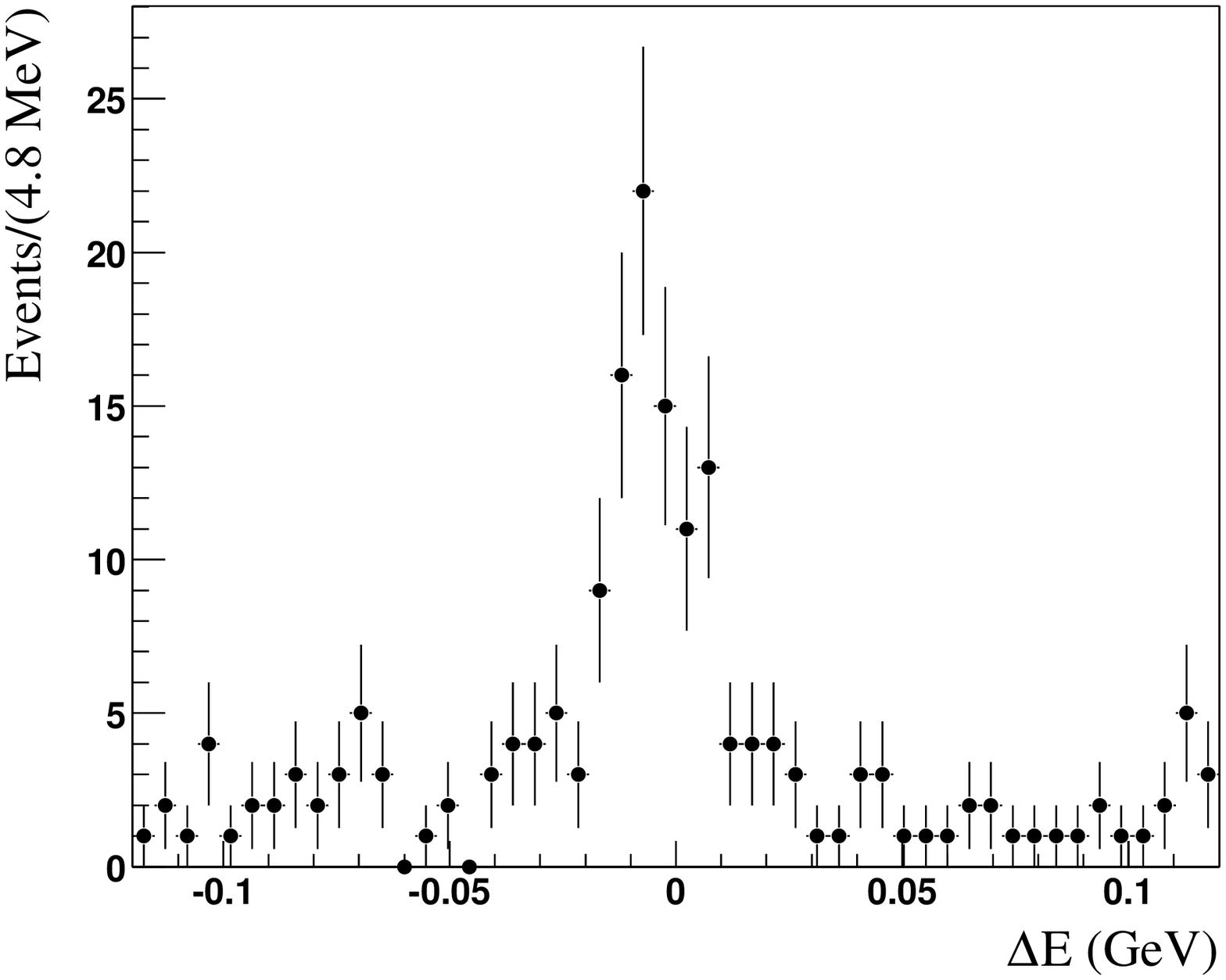}
}

\caption {$\mes$ (left-hand column) distributions from data after applying
requirement $-0.035<\DeltaE<0.03\,\gev$. The $\Delta E$ (right-hand column)
distributions from data after applying requirement $\mes > 5.27 \,\gev/c^2$.
The plots from top to bottom are for $\bds$, $\bd$, $\bdsz$, 
and $\bdz$, respectively.}
\label{fig:mes}
\end{center}
\end{figure}

The signal yield from the ML fit and 
statistical significance for each of the \B modes are
summarized in Table~\ref{tab:effi-yields}. The statistical significance is
calculated as $\sqrt{2\Delta \log{\cal L}}$, where $\Delta \log{\cal L}$ is
the log-likelihood difference between a signal hypothesis corresponding to
the yield and that corresponding to a null yield.
Figure~\ref{fig:proj} shows projection plots of $\mes$ and $\Delta E$ from the
likelihood fit.  The fit shows satisfactory agreement with distributions of
the discriminating variables.

To correct for the efficiency variation across the Dalitz plot, which is of
order 20\% for $\bdsz$ and 5\% for $\bdz$ three-body  
decays, each event is assigned a signal weight, $W_{sig}$, which takes
correlations into account:
\begin{equation}
  W_{sig} = \frac{\sum_j V_{sig,j} {\cal P}(\mes, \Delta E)}{\sum_j N_j
        {\cal P}(\mes, \Delta E)}
\label{eqn:br3}
\end{equation}
where $N_j$ is the number of events, ${\cal P}(\mes, \Delta E)$ is the PDF
of the $j^{th}$ component given the event values of $\mes$ and $\Delta E$, 
and $V_{sig,j}$ is the signal row of the covariance matrix of the components
yields obtained from the likelihood fit. We determine the
efficiency as a function of position on the Dalitz plot from simulated
signal events. The branching fractions of $\bdsz$ and $\bdz$ three-body
decays are corrected by the Dalitz plot dependent efficiencies.  
The effect of the $D^{*-}$ polarization in the $\bdsz$ mode is neglected in
this Dalitz plot dependent efficiency correction.  

We estimate the efficiencies for 4-body decays with the same method as for
three-body decays. We find that the differences are 7.5\% and 3.1\% from the
efficiencies obtained from a pure phase-space MC for the $\bds$ and $\bd$,
respectively.  We will assign 7.5\% and 3.1\% as additional systematic
errors for the two four-body decays, respectively.
 
\begin{figure}[!htb]
\begin{center}
\resizebox{15cm}{!}{
                \includegraphics{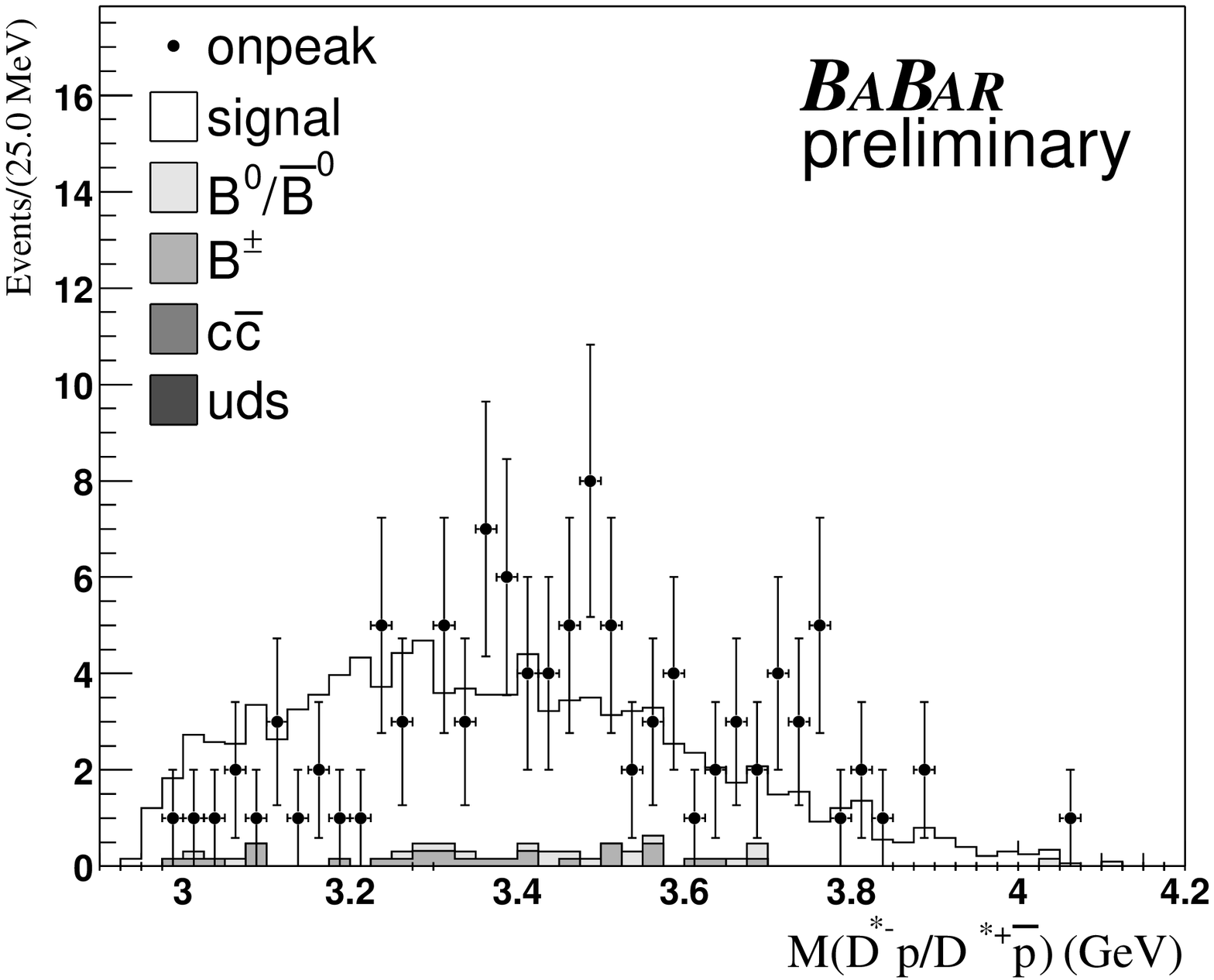}
                \includegraphics{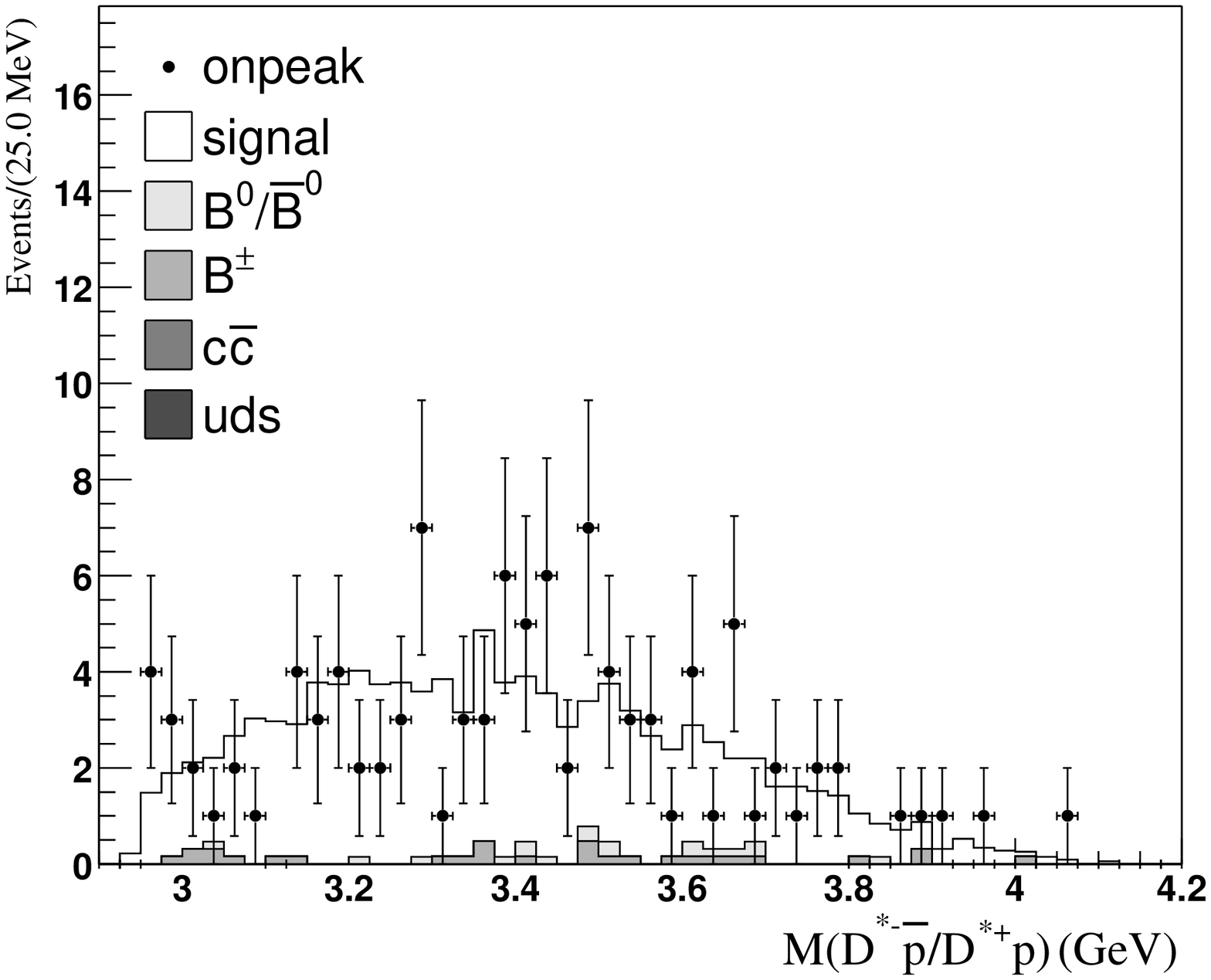}
 }
\caption{The invariant mass distribution of $\Dstarm\proton$ (left) and $\Dstarm\antiproton$ 
(right) combinations from signal candidates in the $\bds$ decay mode. The cross-hatched histograms 
describe the
contributions from different background processes. The open histogram is 
the expected contribution from $\bds$ phase space signal MC simulation.}
\label{fig:ch1_1}
\end{center}
\end{figure}

In each of the four \B decay modes, the signal region is defined as:
$-0.035<\DeltaE<0.03\,\gev$ and $\mes>5.27\,\gev$. The \mes\ distributions
after applying the requirement $-0.035<\DeltaE<0.03\,\gev$, and $\Delta E$
distributions after applying the requirement $\mes > 5.27 \,\gev/c^2$, 
are shown in Fig.~\ref{fig:mes} for all four decay modes. We also compare
the invariant mass spectra for charmed meson and baryon combinations with
a pure phase space hypothesis for \B decays. Figure~\ref{fig:ch1_1} shows the invariant
mass of $\Dstarm\proton$ and $\Dstarm\antiproton$ for signal candidates in the $\bds$ mode. The
open histogram is the expected distribution from $\bds$ phase space signal MC simulation
which is normalized to the fitted signal yield in data.
The cross-hatched histograms describe the contributions from different background
processes, which is normalized to the fitted background yield in data.

\begin{figure}[!htb]
\begin{center}
\resizebox{15cm}{!}{
                \includegraphics{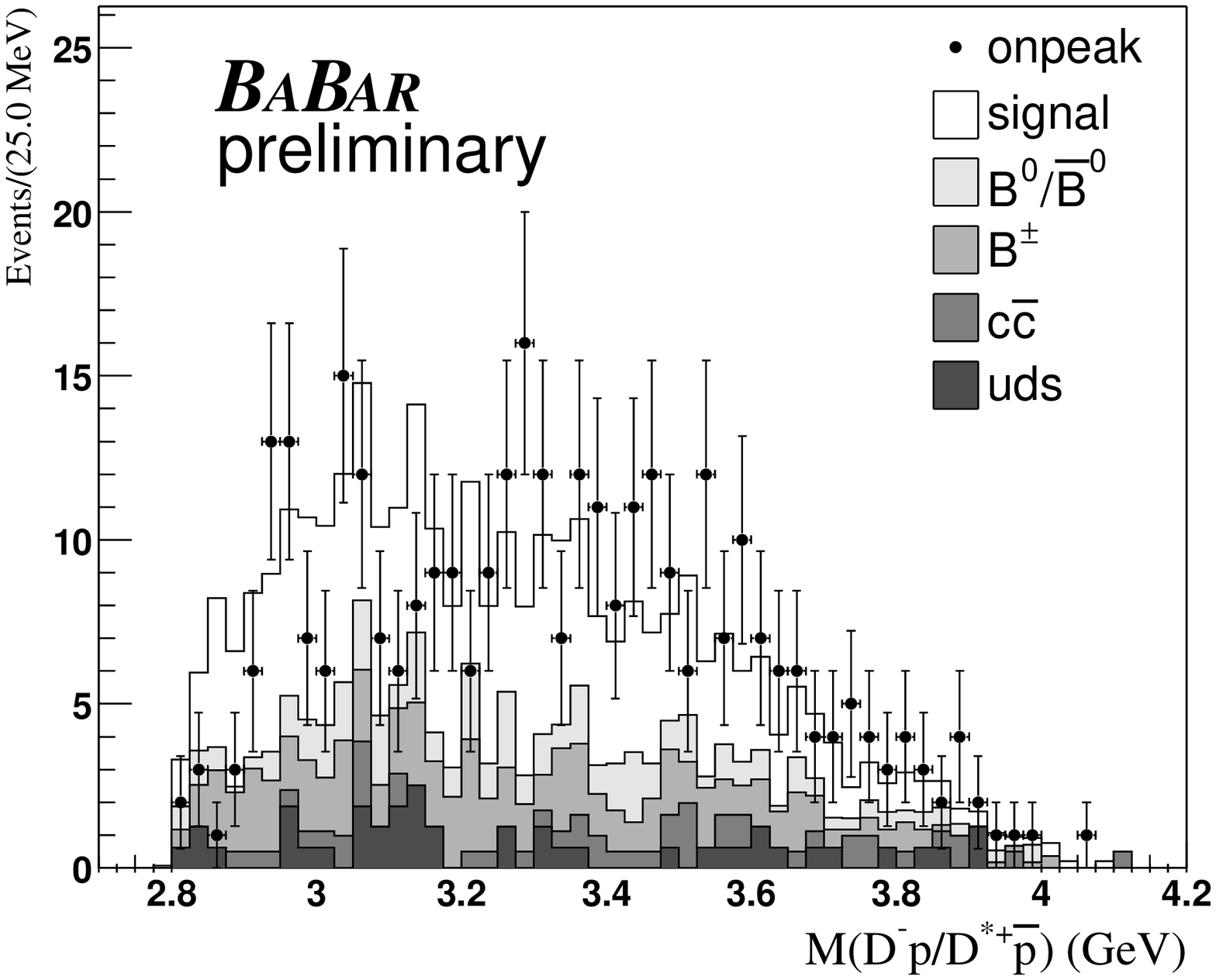}
                \includegraphics{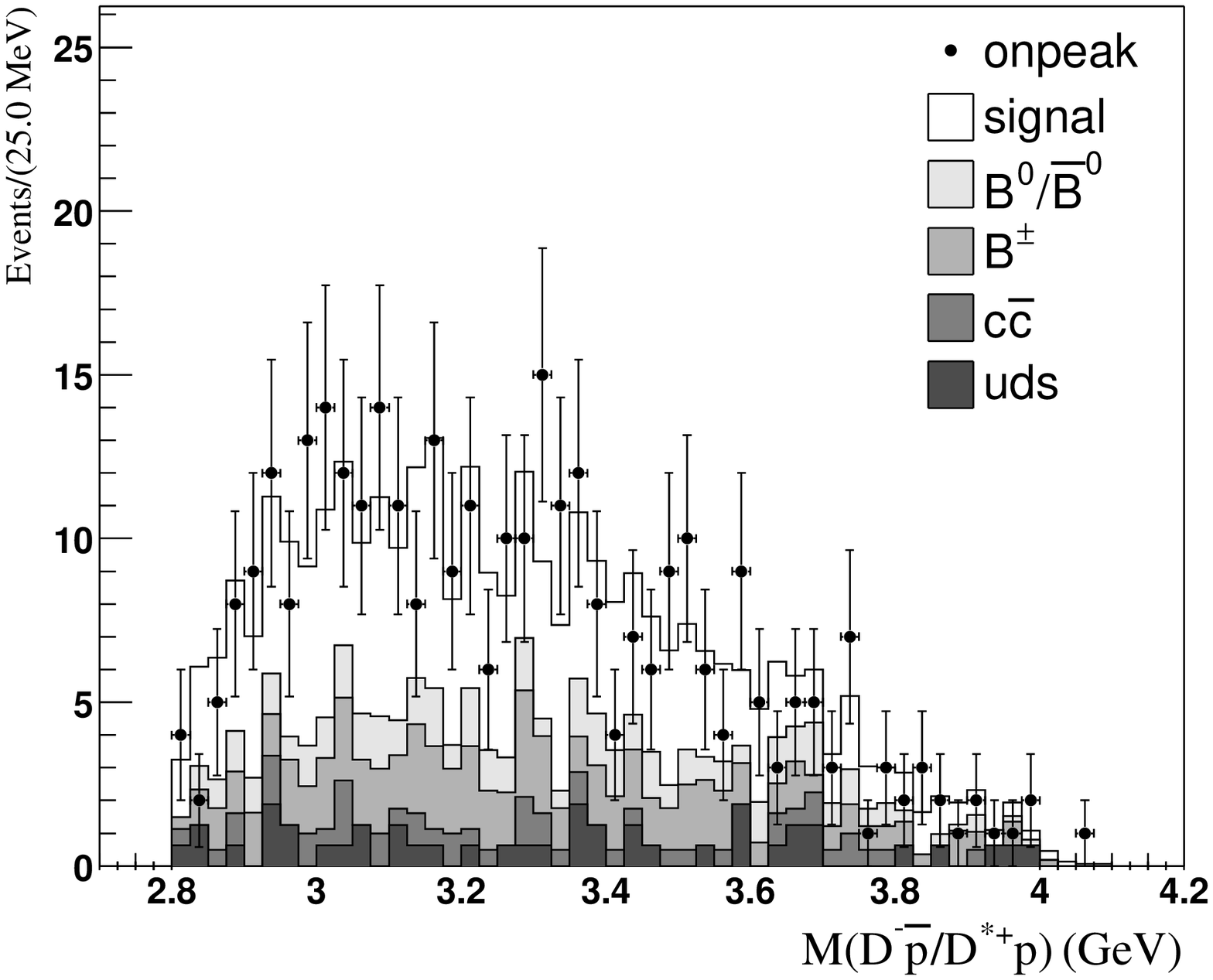}
 }
\caption{The  invariant mass distribution of $\Dm\proton$ (left) and $\Dm\antiproton$ (right) 
combinations from signal candidates in the $\bd$ decay mode. The cross-hatched histograms
describe the
contributions from different background processes. The open histogram is 
the expected contribution from the $\bd$ phase space signal MC simulation.}
\label{fig:ch2_1}
\end{center}
\end{figure}

\begin{figure}[!htb]
\begin{center}
\resizebox{15cm}{!}{
                \includegraphics{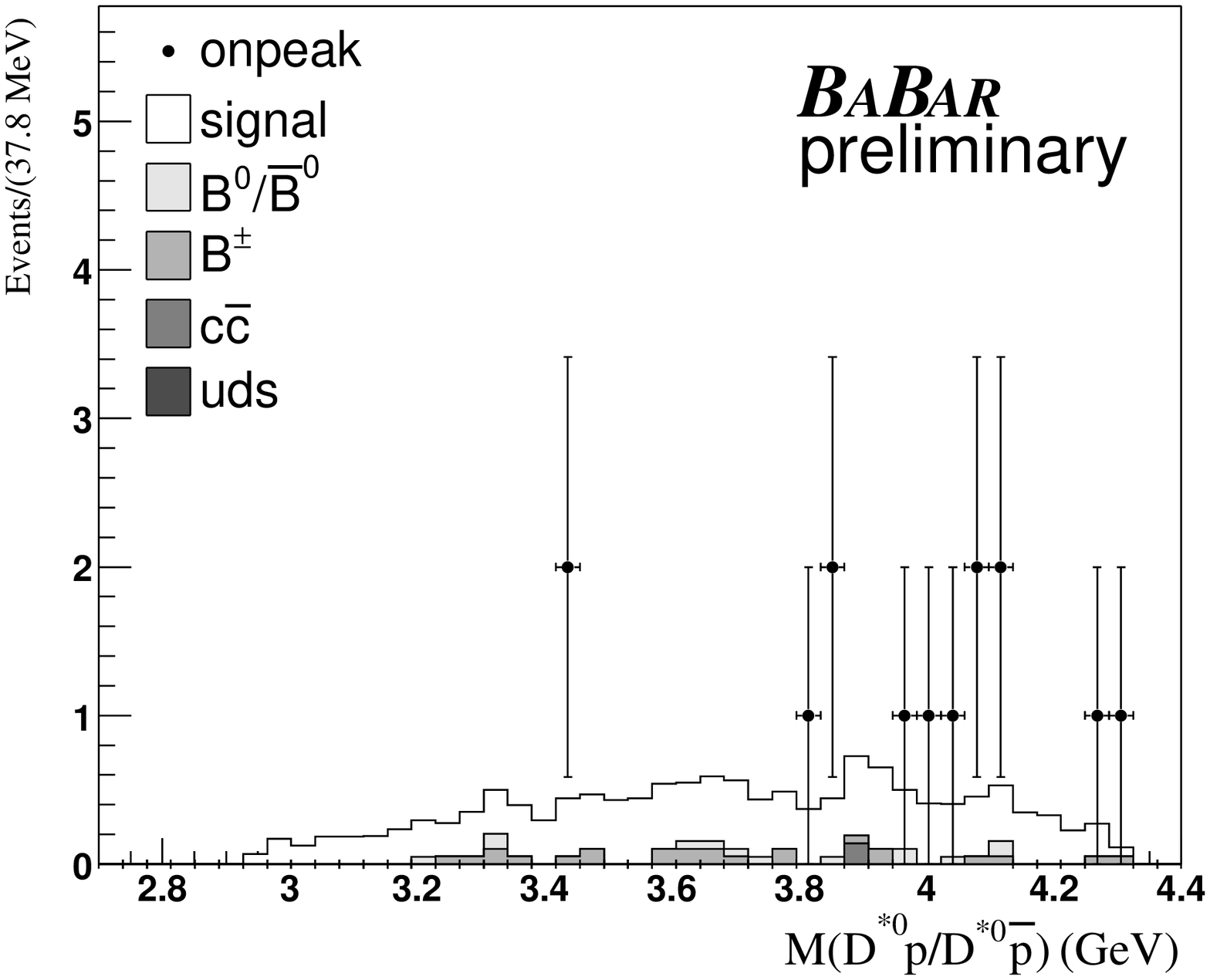}
                \includegraphics{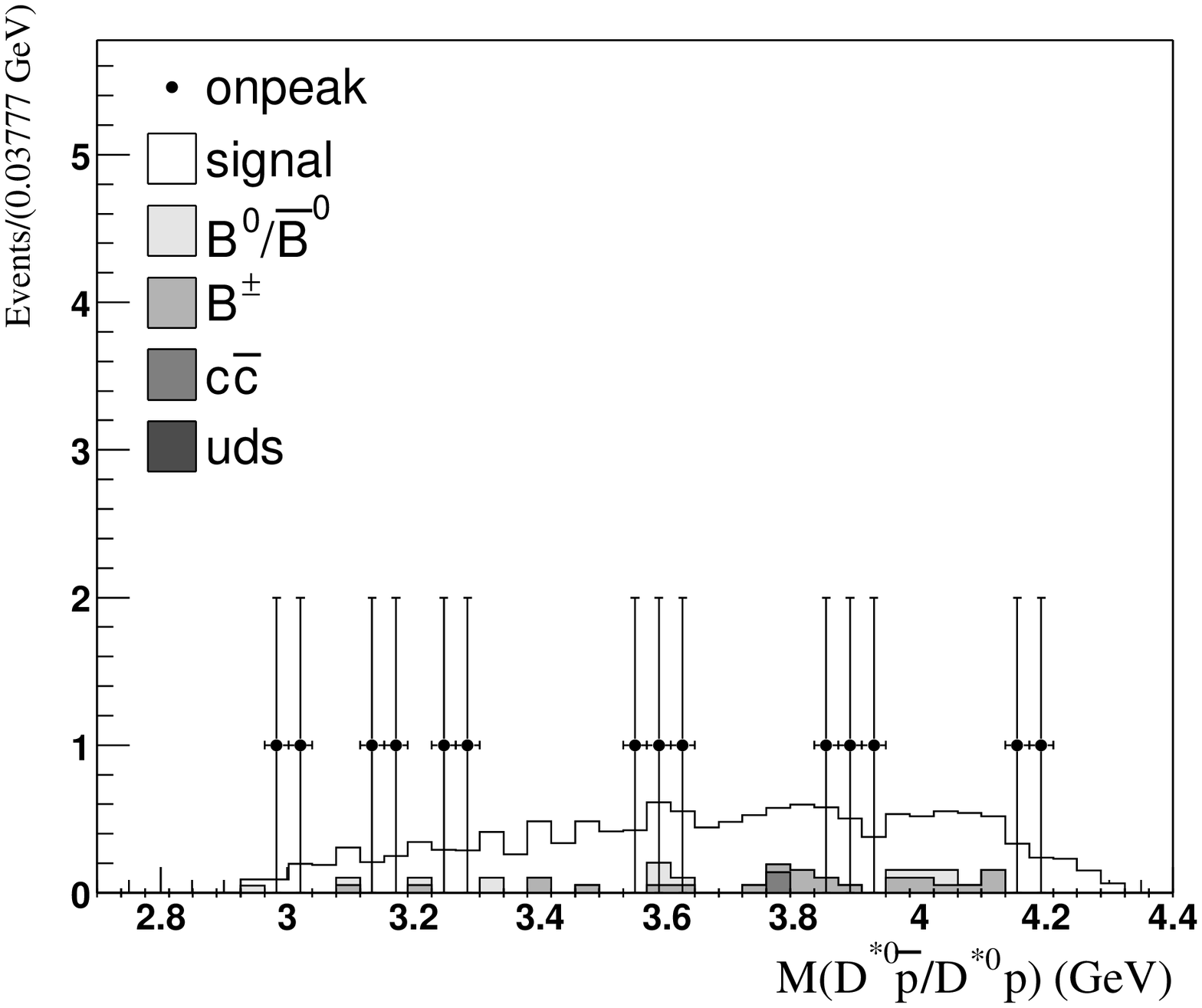}
 }
\caption{The invariant mass distribution of $\Dstarzb\proton$ (left) and 
$\Dstarzb\antiproton$ (right) combinations from signal candidates in the $\bdsz$ decay mode. 
The cross-hatched histograms describe the
contributions from different background processes. The open histogram is
the expected contribution from the $\bdsz$ phase space signal MC simulation.}
\label{fig:ch3_1}
\end{center}
\end{figure}

\begin{figure}[!htb]
\begin{center}
\resizebox{15cm}{!}{
\includegraphics{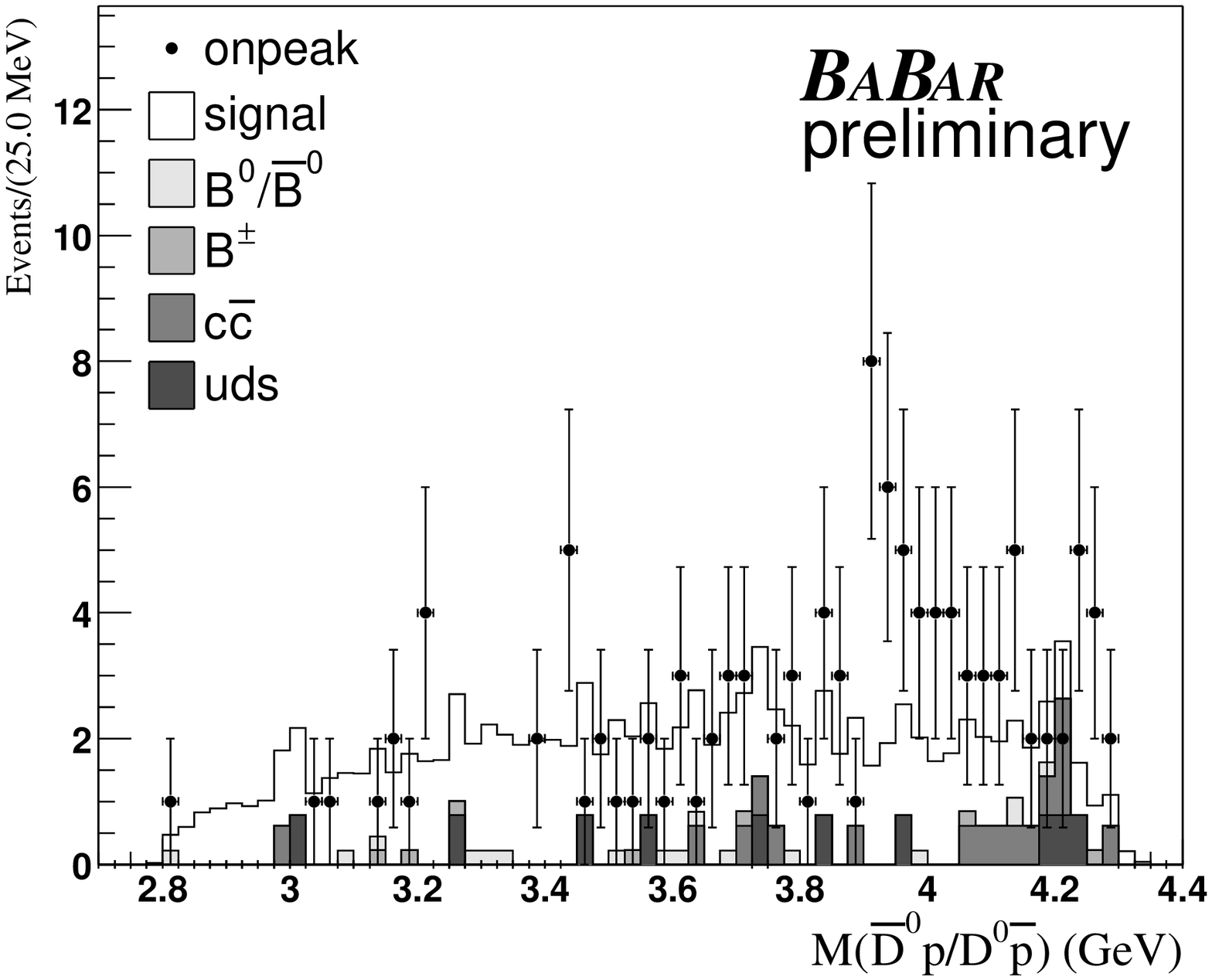}
\includegraphics{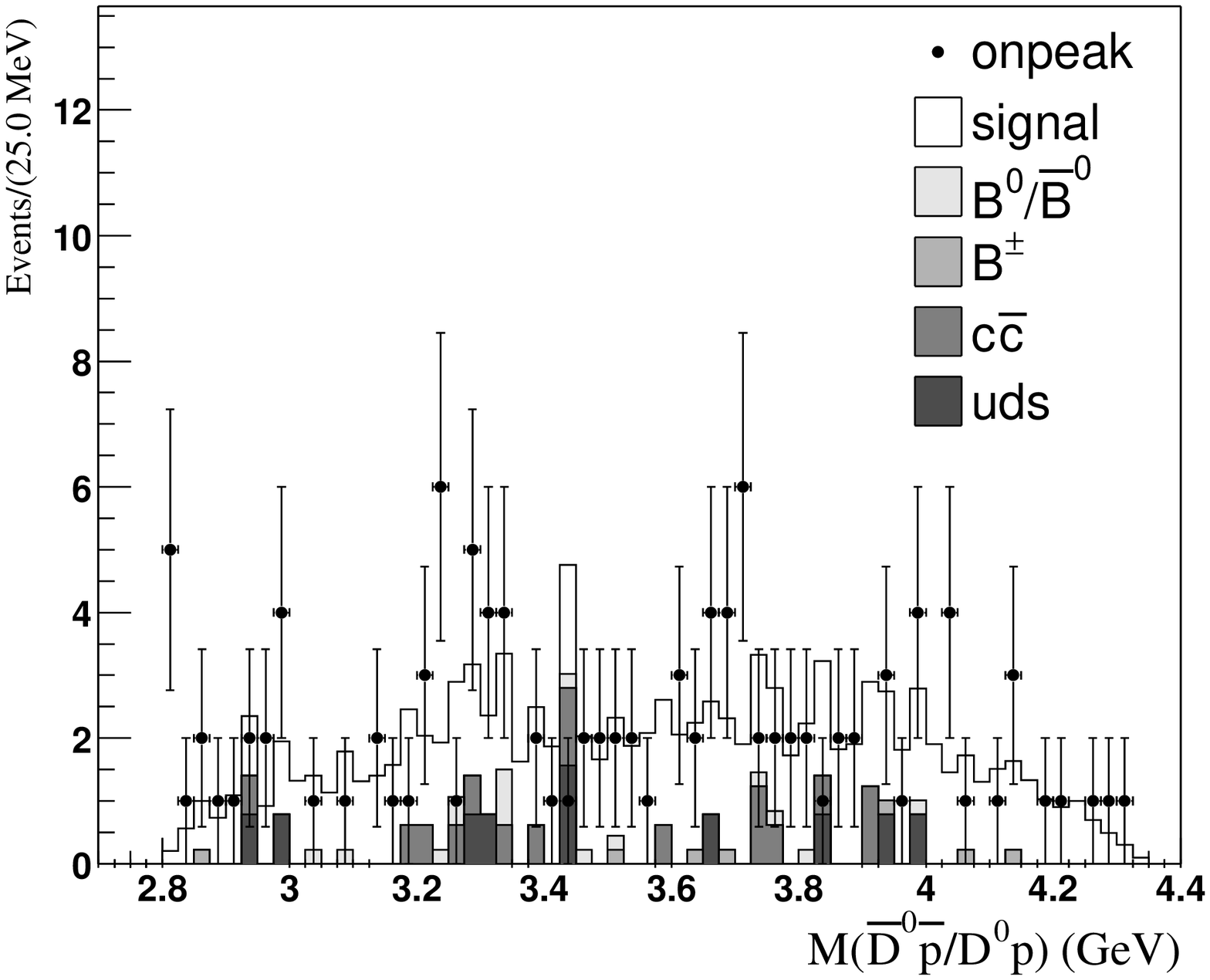}
}
\caption{ The invariant mass distribution of $\Dzb\proton$ (left) and $\Dzb\antiproton$ (right) 
combinations from signal candidates in the $\bdz$ decay mode. The cross-hatched histograms
describe the contributions from different background processes. The open
histogram is the expected contribution from the $\bdz$ phase space signal MC simulation.}
\label{fig:ch4_1}
\end{center}
\end{figure}

\begin{figure}[!htb]
\begin{center}
\includegraphics[height=8cm]{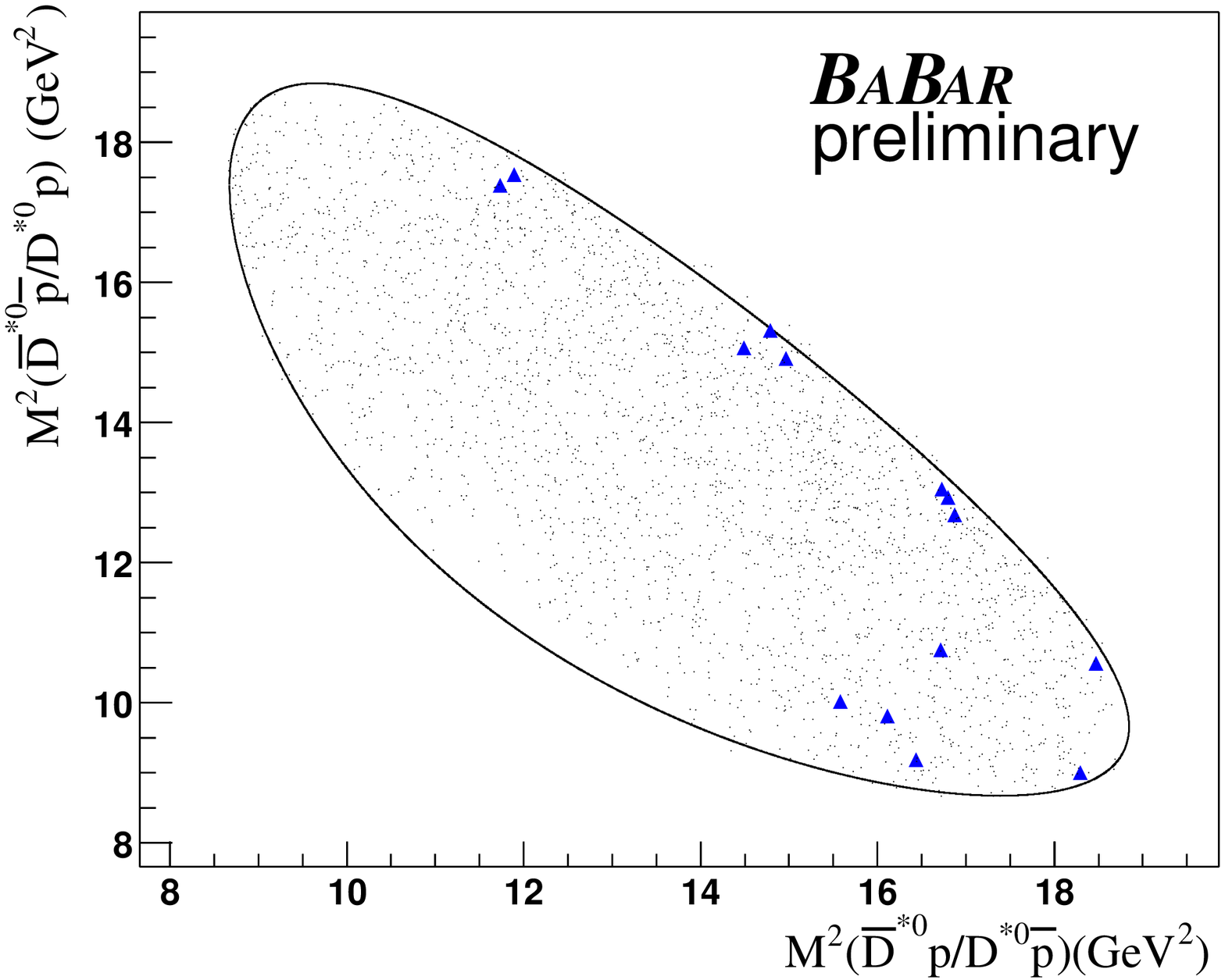}
\caption{Dalitz plot for the mode $\bdsz$. 
The horizontal and
vertical axes correspond to the squared invariant masses of the $\Dstarzb\proton$ and
$\Dstarzb\antiproton$,
respectively. The triangles and dots correspond to signal events in data and the phase space
simulation, respectively.} \label{fig:ch3_2}
\end{center}
\end{figure}

\begin{figure}[!htb]
\begin{center}
\includegraphics[height=8cm]{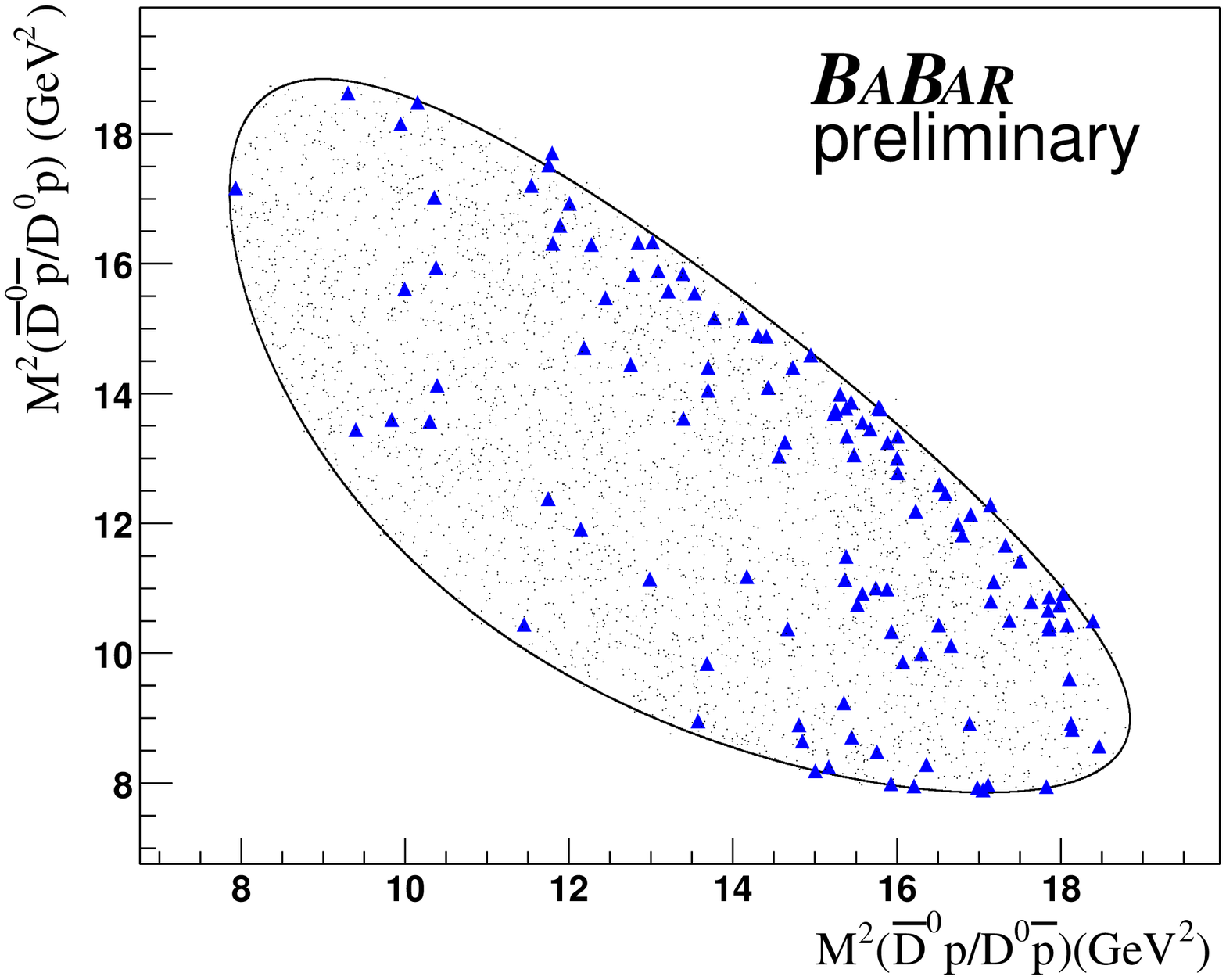}
\caption{Dalitz plot for the mode $\bdz$. 
The horizontal and
vertical axes correspond to the squared invariant masses of the $\Dzb\proton$ and
$\Dzb\antiproton$, respectively. The triangles and dots correspond to signal
events in data and the phase space    
simulation, respectively.}
\label{fig:ch4_2}
\end{center}
\end{figure}
\begin{figure}[bp]
\begin{center}
\resizebox{15cm}{!}{
        \includegraphics{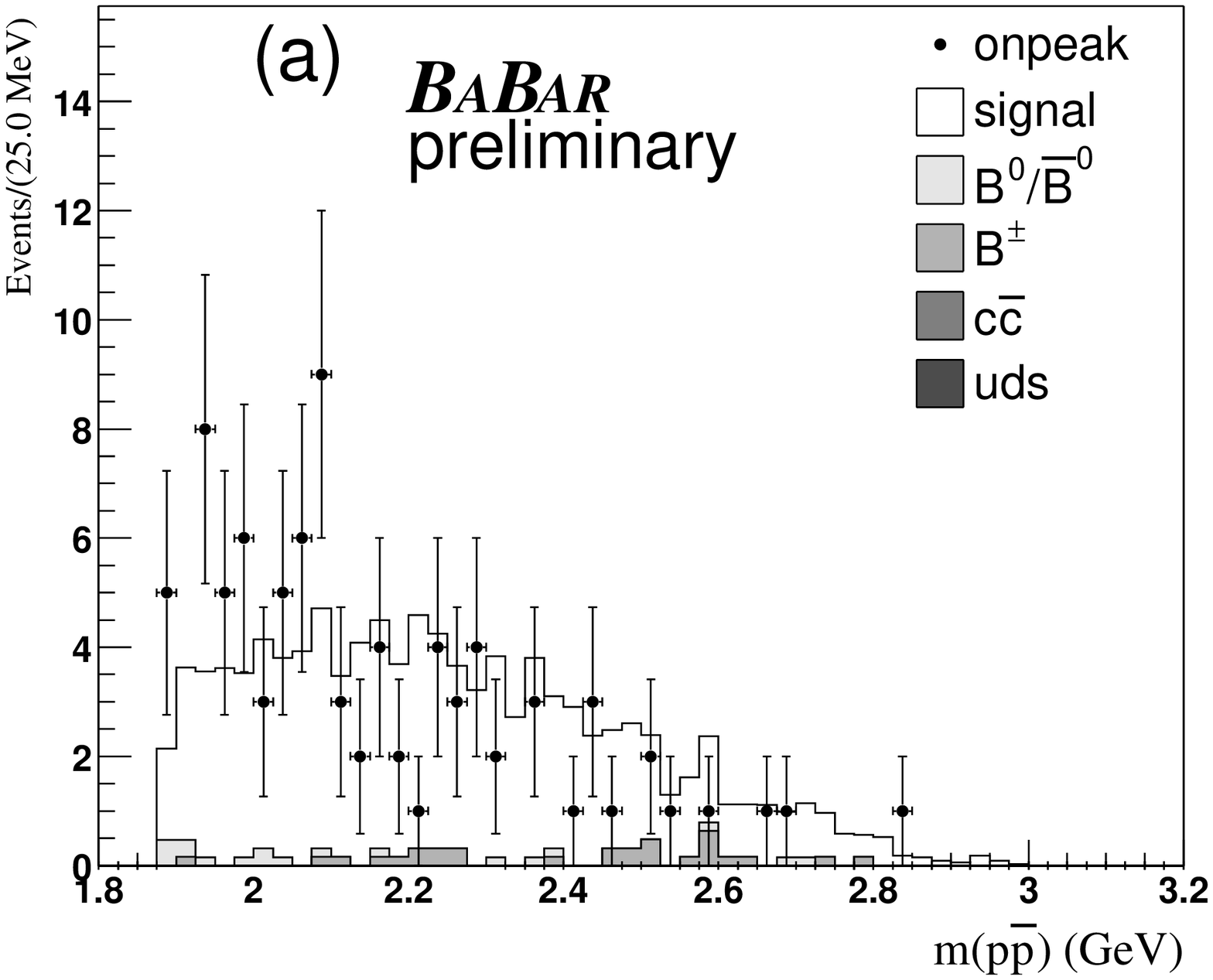}\hspace{0.3cm}
        \includegraphics{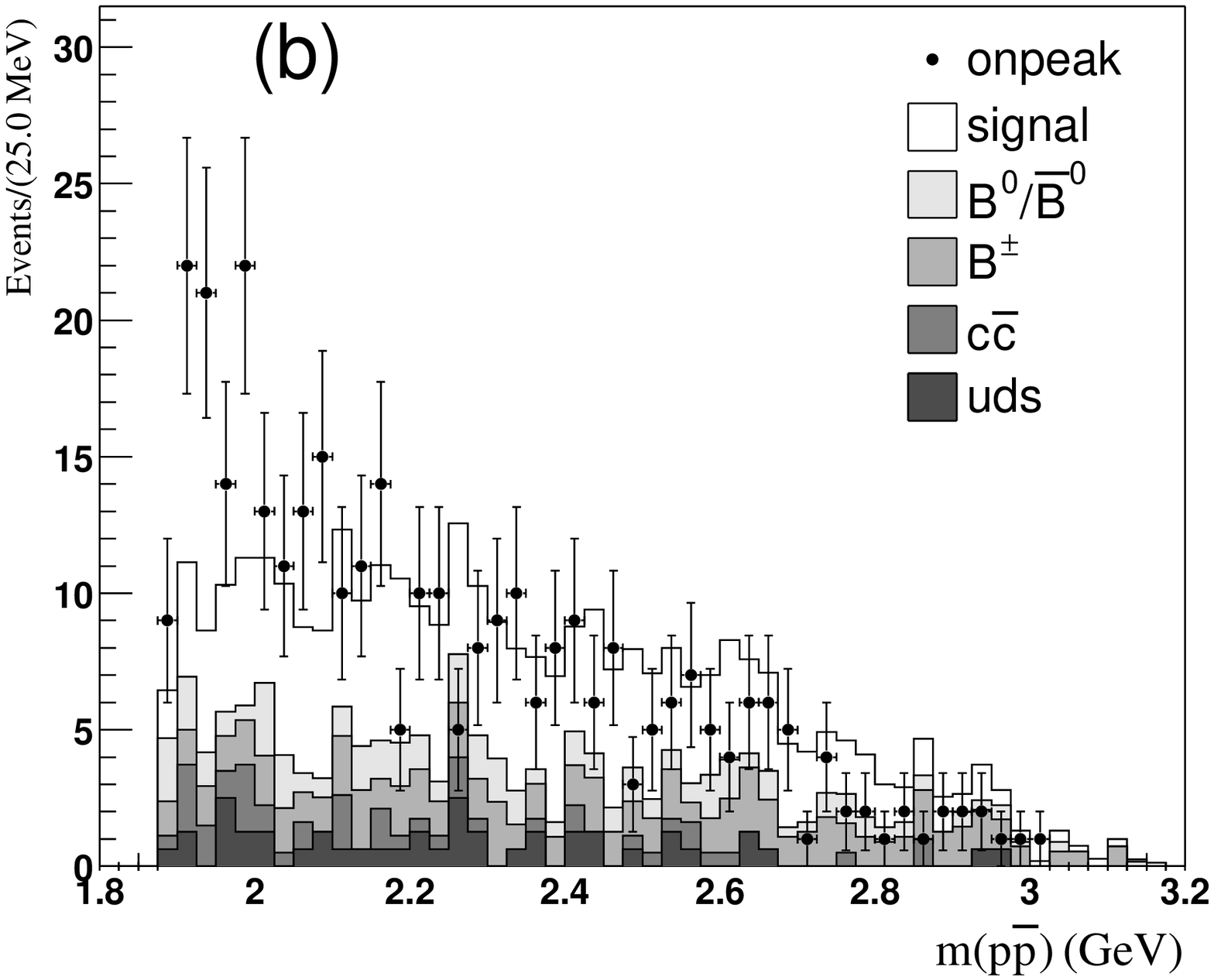}
}
\resizebox{15cm}{!}{
        \includegraphics{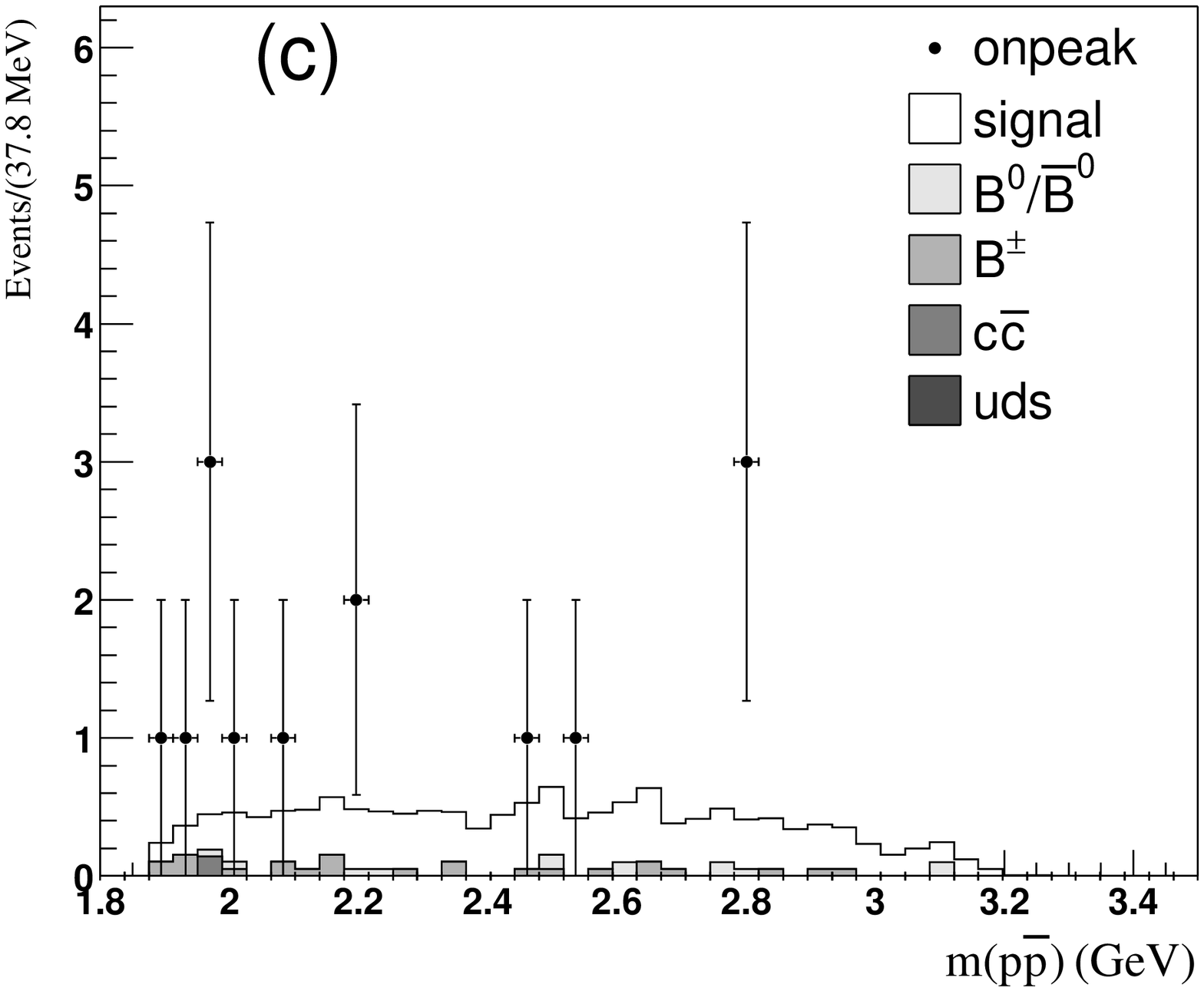}\hspace{0.3cm}
        \includegraphics{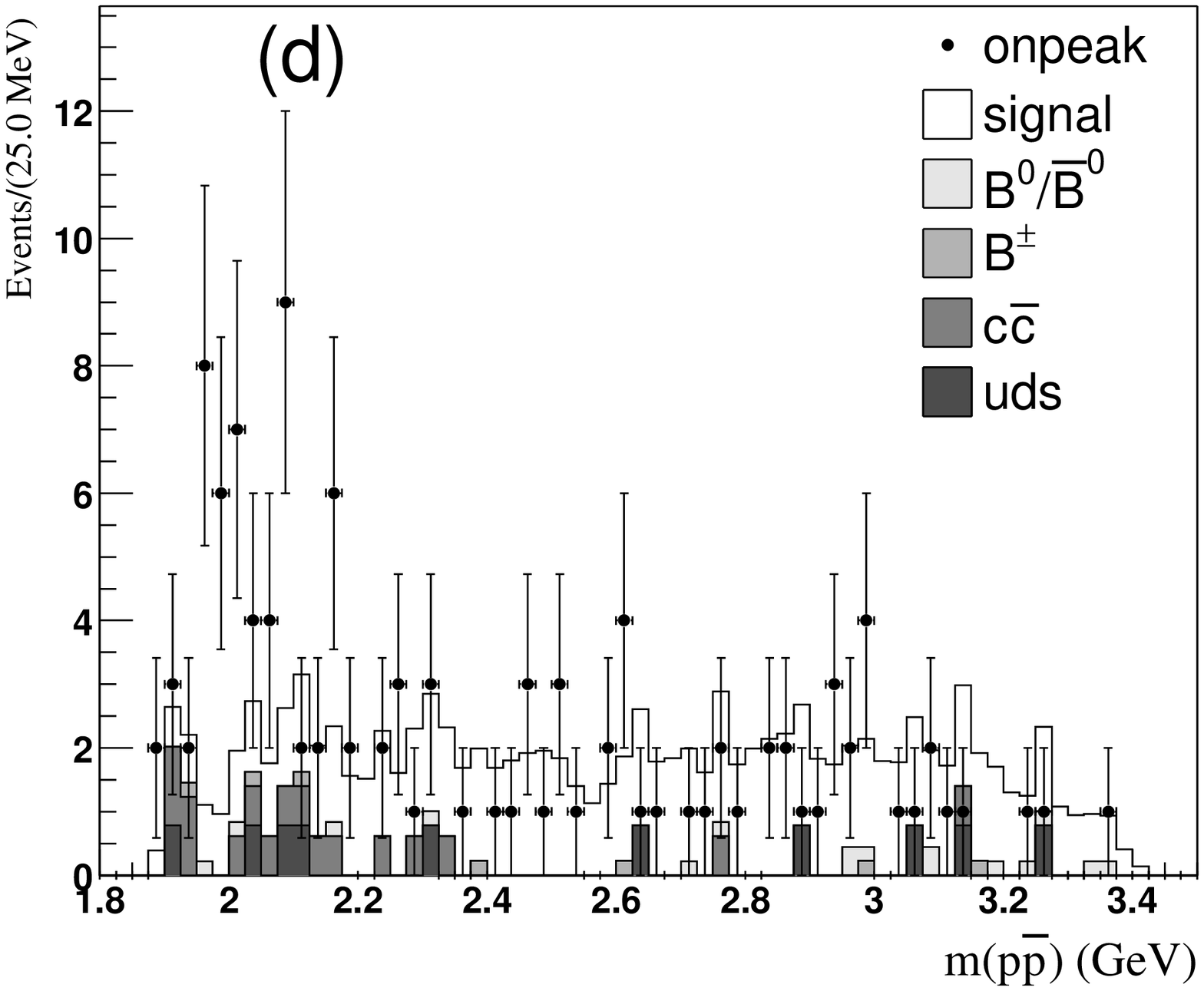}
}
\caption {Invariant mass distributions for $p\antiproton$ pairs in the
signal sample for (a) for $\bds$; (b) for $\bd$; (c) for $\bdsz$; and (d)
for $\bdz$, respectively. 
The cross-hatched histograms describe the contributions from different background processes. 
The open histogram is the expected contribution from the phase space signal MC simulation. }
\label{fig:pp}
\end{center}
\end{figure}

The shape of the invariant mass spectra in Fig.~\ref{fig:ch1_1} show that the 
$\Dstarm\proton$ distribution in data is only marginally consistent with
a phase space model, while the $\Dstarm\antiproton$ is quite consistent
with phase space. A similar observation holds for the $\Dm\proton$ and
$\Dm\antiproton$ distribution from signal candidates in the mode $\bd$ 
as shown in Fig.~\ref{fig:ch2_1}.  

As shown in Fig.~\ref{fig:ch3_1}, the shapes of the invariant mass spectra from
a phase-space model for the $\Dstarzb\proton$ and $\Dstarzb\antiproton$
distributions in the $\bdsz$ mode,
are consistent with data given the current low statistics.
However, the shape for the $\Dzb\proton$ distribution 
in the $\bdz$ mode is inconsistent and it appears to deviate from
the phase-space expectation, as illustrated in the left-hand plot of
Fig.~\ref{fig:ch4_1}.  For $\Dzb\antiproton$, the distribution in data is
consistent with the phase space, as shown in the right-hand plot of Fig.~\ref{fig:ch4_1}.

To further compare the observed mass distributions in data with uniform
phase space, we show the Dalitz plot distributions for the two three-body
decay modes in Figures~\ref{fig:ch3_2} and~\ref{fig:ch4_2}. The distribution of
$m^2(\Dstarzb\proton)$ versus $m^2(\Dstarzb\antiproton)$ for signal
candidates in the $\bdsz$
mode is shown in Figure~\ref{fig:ch3_2}; the distribution between $m^2(\Dzb \proton)$
and $m^2(\Dzb\antiproton)$ for the $\bdz$ mode is shown in Figure~\ref{fig:ch4_2}.
The triangles and dots correspond to data and the phase space MC simulation.
Figure~\ref{fig:ch3_2} shows a Dalitz plot for the mode
$\bdsz$. The comparison with the phase space distribution is limited by low statistics. 
The equivalent plot for the
mode $\bdz$, shown in Figure~\ref{fig:ch4_2}, shows a deviation from the
pure phase space model.  
In addition, the Dalitz plots display a threshold
enhancement in the $\proton\antiproton$ invariant mass spectrum.
Figure~\ref{fig:pp} shows the invariant masses of $p\antiproton$ for each \B
decay mode. A threshold enhancement is observed for each \B decay mode. 
A similar effect has also been observed by other experiments~\cite{prl_89_151802, bes}.
 
The compatibility of data and MC simulation was quantified by means of a
Kolmogorov-Smirnov test. Table~\ref{tab:ks} shows the
probabilities of compatibility of data and MC simulation, where the shape of the data
points and the MC histograms shown in Figures~\ref{fig:ch1_1}-\ref{fig:ch4_1}
and~\ref{fig:pp} were compared. 
The first and second columns
correspond to the invariant mass spectra of the relevant $D$-meson and
$\proton$ combination (the corresponding minimal quark content is given in parentheses). 
Charge conjugation is assumed. The last column corresponds to the invariant mass spectra of 
the $\proton\antiproton$ system. 
We draw a number of observations from the calculated probabilities in
Table~\ref{tab:ks}. For the
first column, the top two rows are marginally consistent with phase
space, while the third is consistent, and the last row is inconsistent.
The entire second column is consistent with phase space, while in the third column only the
third row is consistent, and all others are inconsistent.

\begin{table}[!htb]
\caption{Probabilities of obtaining the observed result under the null
hypothesis that the data is sampled from the phase space model.  
}
\begin{center}   
\begin{tabular}{c|c|c|c} \hline
&  &  &    \\    
$\bds$ &  $\Dstarm\proton (\cbar\d\u\u\d)$ & $\Dstarm\antiproton
(\cbar\d\ubar\ubar\dbar)$ & M($\proton\antiproton$) \\
&       1.32$\times 10^{-2}$ & 0.83 & 1.22$\times 10^{-4}$ \\ \hline
&  &  &    \\    
$\bd$ &  $\Dm\proton (\cbar\d\u\u\d)$ & $\Dm\antiproton
(\cbar\d\ubar\ubar\dbar)$ & M($\proton\antiproton$) \\
&       1.78$\times 10^{-2}$ & 1.00 & 2.58$\times 10^{-4}$ \\ \hline
&  &  &    \\    
$\bdsz$ & $\Dstarzb\proton (\cbar\u\u\u\d)$ &
$\Dstarzb\antiproton (\cbar\u\ubar\ubar\dbar)$ & M($\proton\antiproton$) \\
&       0.13 & 0.68 & 0.32 \\ \hline
&  &  &    \\    
$\bdz$ & $\Dzb\proton (\cbar\u\u\u\d)$ &
$\Dzb\antiproton (\cbar\u\ubar\ubar\dbar)$ & M($\proton\antiproton$) \\
&       3.51$\times 10^{-4}$ & 0.77 & 3.26$\times 10^{-4}$ \\ \hline
\end{tabular}    
\end{center}     
\label{tab:ks}   
\end{table}


\section{SYSTEMATIC STUDIES}
\label{sec:Systematics}

\begin{table}[h]
\begin{center}
\caption{ Summary of the systematic uncertainties for all \B modes in
this report.}
\label{tab:sys}
\begin{tabular}{lcccc}
\hline\hline
              & $\Dstppbarpi$ & $\Dcppbarpi$ & $\Dszppbar$ & $\Dzppbar$ \\ 
 Error source & \multicolumn{4}{c}{(in \%)} \\ \hline
 Event selection        & 3.1 & 3.9 & 5.6 & 7.6 \\
 Signal model           & 3.0 & 3.2 & 9.0 & 0.5  \\
 Track finding          & 9.2 & 8.4 & 4.8 & 4.8 \\
 $\pi^0$ reconstruction & - & - & 5.6  & - \\
 Particle ID              & 0.1 & 0.3 & 0.1 & 0.1 \\
 Phase space MC         & 7.5 & 3.1 &- & - \\
 Number of \B mesons     & 1.1 &1.1  &1.1  &1.1 \\
 Daughter branching fraction & 2.5 & 6.5 & 5.3 & 2.4  \\ \hline
 Total                       & 13.0 & 12.2 & 13.9 & 9.4  \\ \hline \hline
\end{tabular}
\end{center}
\end{table}
                                                 
Systematic uncertainties in the ML fit originate from the modeling of the
PDFs.  We vary the PDF parameters within their respective uncertainties,
and derive the associated systematic errors. 
The SCF fractions can be floated in the fit to data for the decays
$\BtoDstppbarpi$ and $\bd$. The
differences between MC and data are 8\% and 16\% for the two decays, respectively.
For the systematic uncertainty due to the SCF fraction, we vary the 
SCF fraction by $\pm 10\%$ in the fit according to the above validation from
data. We also perform fits to large MC samples with signal and combinatorial
background. No bias is observed in these tests.

The systematic errors in the efficiencies are for the track finding (1.2\% -
1.4\% per track and 2.2\% for the slow charged $\pi$), particle identification 
(0.1 - 0.3 \%) and $\pi^0$ reconstruction (5.6\%). The systematic error in
each \B decay mode arising from variations of the selection criteria 
is also shown in Table~\ref{tab:sys}. 
The reconstruction efficiencies depend on the final state distributions due to \B decay
dynamics. As discussed earlier, we assign 7.5\% and 3.1\% systematic uncertainties to
account for the uncertainty on the efficiencies from the phase-space MC
simulation for $\BtoDstppbarpi$ and $\bd$ four-body decays, respectively.
Other systematic effects are from event-selection criteria, daughter
branching fractions~\cite{ref:pdg2002}, 
MC statistics, and the number of \B mesons in the sample~\cite{bib:BCount}. 
The contributions to the systematic errors on signal yields are summarized in
Table~\ref{tab:sys}.


\section{SUMMARY}
\label{sec:Summary}

The $\Bz$ meson decay modes $\bdsshort\bdshort\bdszshort$, and $\bdz$
have been studied in a data sample equivalent 113 fb$^{-1}$ of integrated luminosity.  
Table~\ref{tab:branching} summarizes the branching fractions obtained for the
decays $\BtoDstppbarpi$, $\BtoDcppbarpi$, $\BtoDszppbar$ and
$\BtoDzppbar$.
These results are compared with previous measurements performed by the
CLEO~\cite{prl_86_2732} and Belle~\cite{prl_89_151802}
collaborations, when possible, and good agreement is found.
The $\bd$ decay has been observed for the first time with the 
branching fraction $\BR(\bd)=(3.80 \pm 0.35\pm 0.46)\times 10^{-4}$. 
All results are preliminary.

\begin{table}[!htb]
\caption{The branching ratios (in units of $10^{-4})$ for the $\Bz$
modes are measured here. Statistical (first) and systematic
(second) errors are given. The significance includes both the statistical
error and the systematic error.
Results obtained here are compared with
results published by the CLEO~\cite{prl_86_2732} and
Belle~\cite{prl_89_151802} collaborations (when available).}
\begin{center}
\begin{tabular}{cccc} \hline
Final state & $\BR (10^{-4})$ & Significance & Reference $\BR (10^{-4})$ \\
\hline \hline
$\bds$ & $5.61\pm 0.59\pm 0.73$ & 13$\sigma$ & $6.5^{+1.3}_{-1.2}\pm 1.0$~\cite{prl_86_2732}\\
$\bd $ &  $3.80\pm 0.35\pm 0.46$ &13$\sigma$ & - \\
$\bdsz$  &  $0.67\pm 0.21\pm 0.09$ & 4.0$\sigma$  & $1.20^{+0.33}_{-0.29}\pm 0.21$~\cite{prl_89_151802}\\
$\bdz$  &  $1.24\pm 0.14\pm 0.12$ & 12$\sigma$ & $1.18 \pm 0.15 \pm0.16$~\cite{prl_89_151802} \\
\hline
\end{tabular}
\end{center}
\label{tab:branching}
\end{table}

\section{ACKNOWLEDGMENTS}
\label{sec:Acknowledgments}

We are grateful for the 
extraordinary contributions of our \pep2\ colleagues in
achieving the excellent luminosity and machine conditions
that have made this work possible.
The success of this project also relies critically on the 
expertise and dedication of the computing organizations that 
support \babar.
The collaborating institutions wish to thank 
SLAC for its support and the kind hospitality extended to them. 
This work is supported by the
US Department of Energy
and National Science Foundation, the
Natural Sciences and Engineering Research Council (Canada),
Institute of High Energy Physics (China), the
Commissariat \`a l'Energie Atomique and
Institut National de Physique Nucl\'eaire et de Physique des Particules
(France), the
Bundesministerium f\"ur Bildung und Forschung and
Deutsche Forschungsgemeinschaft
(Germany), the
Istituto Nazionale di Fisica Nucleare (Italy),
the Foundation for Fundamental Research on Matter (The Netherlands),
the Research Council of Norway, the
Ministry of Science and Technology of the Russian Federation, and the
Particle Physics and Astronomy Research Council (United Kingdom). 
Individuals have received support from 
CONACyT (Mexico),
the A. P. Sloan Foundation, 
the Research Corporation,
and the Alexander von Humboldt Foundation.

\end{document}